\documentclass[reprint, aps, prb, nofootinbib]{revtex4-2}

\usepackage{xcolor}
\usepackage{graphicx}
\usepackage{color}
\usepackage{amsthm, amssymb, amsmath}
\usepackage{amsfonts}
\usepackage{mathtools}
\usepackage{bm}
\usepackage{physics}
\usepackage[normalem]{ulem}
\usepackage{ascmac}
\usepackage{comment}
\usepackage{empheq}
\usepackage[thicklines]{cancel}
\usepackage{enumerate}
\usepackage{cases}
\usepackage{mathrsfs}
\usepackage{booktabs}
\usepackage{multirow}
\usepackage[version=4]{mhchem}
\usepackage[caption=false, position=top]{subfig}
\allowdisplaybreaks

\newcommand{\up}{\uparrow}
\newcommand{\down}{\downarrow}

\newcommand{\Z}{\mathbb{Z}}

\newcommand{\kx}{k_\mathrm{x}}
\newcommand{\ky}{k_\mathrm{y}}
\newcommand{\kz}{k_\mathrm{z}}

\newcommand{\kb}{k_\mathrm{B}}

\newcommand{\Tc}{T_\mathrm{c}}

\newcommand{\sx}{s_\mathrm{x}}
\newcommand{\sy}{s_\mathrm{y}}
\newcommand{\sz}{s_\mathrm{z}}

\newcommand{\rhox}{\rho_\mathrm{x}}
\newcommand{\rhoy}{\rho_\mathrm{y}}
\newcommand{\rhoz}{\rho_\mathrm{z}}

\newcommand{\sigmax}{\sigma_\mathrm{x}}
\newcommand{\sigmay}{\sigma_\mathrm{y}}
\newcommand{\sigmaz}{\sigma_\mathrm{z}}

\newcommand{\taux}{\tau_\mathrm{x}}
\newcommand{\tauy}{\tau_\mathrm{y}}
\newcommand{\tauz}{\tau_\mathrm{z}}

\newcommand{\M}{\bar{\mathrm{M}}}
\newcommand{\X}{\bar{\mathrm{X}}}
\newcommand{\Y}{\bar{\mathrm{Y}}}

\newcommand{\BdG}{\mathrm{BdG}}
\renewcommand{\wp}{\mathrm{wp}}

\newcommand{\FS}{\mathrm{FS}}

\renewcommand{\P}{\mathcal{P}}

\usepackage{hyperref}
\hypersetup{
    citecolor = blue,
    colorlinks = true,
    urlcolor = blue,
    linkcolor = blue
}

\usepackage{orcidlink}

\begin{document}

\title{Competition between on-site and off-site pairing phases and surface spectra in superconducting wallpaper fermion systems}
\author{Kaito Yoda \orcidlink{0009-0006-5607-6364}}
\email{yoda.kaito.c1@s.mail.nagoya-u.ac.jp}
\affiliation{Department of Physics, Nagoya University, Nagoya 464-8602, Japan}
\author{Ai Yamakage \orcidlink{0000-0003-4052-774X}}
\affiliation{Department of Physics, Nagoya University, Nagoya 464-8602, Japan}

\date{\today}

\begin{abstract}
    We theoretically investigate the competition between on-site and off-site pairings in superconducting topological crystalline insulators hosting wallpaper fermions, a class of surface states protected by nonsymmorphic symmetries.
    In-plane nearest-neighbor pair hopping stabilizes the $\mathrm{A_{1u}}$ phase, while off-site pairing channels yield additional dominant phases with $\mathrm{B_{2u}}$ and $\mathrm{E_u}$ representations.
    We further show that the $\mathrm{B_{2u}}$ pairing state supports the hybridization between wallpaper-fermion spectra themselves, whereas a representative time-reversal-invariant nematic $\mathrm{E_u}$ state exhibits spectral hybridization between wallpaper fermion and double Majorana Kramers pairs, yielding twisted surface energy spectra.
    These results demonstrate that the twisted dispersions arising from hybridization phenomena persist in the presence of off-site pairing interactions.
\end{abstract}

\maketitle

\section{Introduction} \label{sec:introduction}

The surface of topological superconductors provides a fertile platform for realizing unconventional quasiparticles~\cite{qi2011topological, mizushima2016symmetry, sato2016majorana, chiu2016classification, sato-ando2017topological, aguado2017majorana, tanaka2024theory}.
Such boundary excitations can originate from the nontrivial topology of the bulk Bogoliubov--de Gennes (BdG) Hamiltonian, characterized by fundamental symmetries such as time-reversal, particle-hole, and chiral symmetries~\cite{schnyder2008classification, kitaev2009periodic, ryu2010topological}.
A paradigmatic example of exotic quasiparticles is a Majorana fermion, whose non-Abelian properties make it a promising building block for fault-tolerant quantum computations~\cite{ivanov2001non-abelian, kitaev2006anyons, nayak2008non-abelian, alicea2012new, beenakker2013search}.
Superconducting topological materials, including topological insulators~\cite{yonezawa2018nematic}, topological crystalline insulators~\cite{sasaki2012odd, novak2013unusual, he2013full, hashimoto2015surface, kawakami2018topological}, and topological semimetals~\cite{kobayashi2015topological, lu2015crossed, aggarwal2016unconventional, wang2016observation, wang2017discovery}, offer a complementary route to unconventional surface quasiparticles.
For particular pairing symmetries, normal-state-derived surface bands remain gapless and hybridize with superconductivity-induced boundary modes.
In superconducting topological insulators such as $\mathrm{Cu}_x\ce{Bi2}\ce{Se3}$, for example, mirror-odd pairing leaves the surface Dirac fermions ungapped; hybridization of their branches with those emanating from Majorana Kramers pairs pinned at the time-reversal-invariant momentum produces a characteristic twisted surface BdG spectrum~\cite{hao2011surface, hsieh2012Majorana, yamakage2012theory, yamakage2013theory}.

Motivated by this perspective, we previously extended this analysis to systems with nonsymmorphic crystalline symmetry and in particular, studied the surface states of bulk superconductivity in a topological nonsymmorphic crystalline insulator hosting wallpaper fermions~\cite{yoda2026superconducting, yoda2026double}.
Wallpaper fermions are topologically protected surface quasiparticles associated with time-reversal symmetry and two orthogonal glide symmetries; unlike conventional Dirac fermions, they exhibit a fourfold degeneracy at the time-reversal-invariant momentum~\cite{wieder2018wallpaper, ryu2020wallpaper, zhou2021glide, mao2022third, hwang2023magnetic, mizuno2023hall, mizuno2025magnon, yoda2026superconducting, yoda2026double}.
In Ref.~\onlinecite{yoda2026double}, we found that within the tight-binding model with space group $P4/mbm$ symmetry, crystalline symmetry and Fermi--Dirac statistics allow four symmetry-distinct momentum-independent on-site pair potentials, belonging to the $\mathrm{A_{1g}}$, $\mathrm{A_{2g}}$, $\mathrm{A_{1u}}$, and $\mathrm{A_{2u}}$ representations.
A key result was that, among these superconducting pairings, only the $\mathrm{A_{1u}}$ representation supports the coexistence of gapless wallpaper fermions and double Majorana Kramers pairs, whose hybridization yields a characteristic twisted surface spectrum.

This previous result, however, does not determine the microscopic conditions under which the on-site $\mathrm{A_{1u}}$ pairing  is stabilized.
Moreover, electron--electron interactions in realistic systems are not necessarily restricted to the on-site channels and may contain off-site pairing components.
It is therefore important to establish whether the $\mathrm{A_{1u}}$ phase remains stable against the competing off-site pairing channels or is suppressed by them.
When off-site pairing states become dominant, their surface spectra must be clarified to determine whether the hybridization between gapless wallpaper fermions and Majorana Kramers pairs occurs.

In this paper, we address these questions by investigating the phase diagram and surface spectra in superconducting wallpaper fermion systems, including both on-site and off-site pairing channels.
We first examine the competition among the on-site pairings and find that, for the parameter set considered in Ref.~\onlinecite{yoda2026double}, the $\mathrm{A_{1u}}$ state is favored when the in-plane nearest-neighbor pair-hopping amplitude between on-site Cooper pairs is positive, $V_\parallel>0$.
We then incorporate off-site interactions and show that they can stabilize the off-site $\mathrm{B_{2u}}$ and $\mathrm{E_u}$ phases.
In particular, part of the $\mathrm{A_{1u}}$ pairing phase is replaced by the $\mathrm{E_u}$ pairing phase.
We further analyze the crystalline-symmetry-enforced feature of the surface spectra for the off-site $\mathrm{B_{2u}}$ pairing state and a representative time-reversal-invariant nematic $\mathrm{E_u}$ state.
Both the $\mathrm{B_{2u}}$ state and the nematic $\mathrm{E_u}$ state considered here exhibit characteristic twisted surface dispersions arising from distinct hybridization mechanisms: between gapless wallpaper fermions themselves in the former and between wallpaper fermions and double Majorana Kramers pairs in the latter.
Thus, although the off-site interactions stabilize the $\mathrm{B_{2u}}$ and $\mathrm{E_u}$ phases, these pairing states also support twisted surface dispersions, indicating that hybridization phenomena can persist in the presence of off-site pairing.

The paper is organized as follows.
In Sec.~\ref{sec:model}, we introduce the lattice model for wallpaper fermions~\cite{wieder2018wallpaper} and formulate the on-site and off-site pairing interactions.
In Sec.~\ref{sec:phase_diagram}, we derive the linearized gap equation and present the superconducting phase diagrams.
In Sec.~\ref{sec:surface_state}, we examine the surface spectra for off-site $\mathrm{B_{2u}}$ and $\mathrm{E_u}$ pairings.
In Sec.~\ref{sec:discussion}, we discuss the relation between the interaction parameters and the corresponding surface spectral structures.
Finally, Sec.~\ref{sec:conclusion} summarizes this paper.

\section{Model} \label{sec:model}

Wallpaper fermions are surface states protected by $\Z_4\times\Z_2$ topological invariants in a time-reversal-invariant system with two orthogonal glide symmetries~\cite{wieder2018wallpaper}.
The wallpaper groups that support such surface states are $pgg$ and $p4g$; in this work, we consider the higher-symmetry case, where the surface is invariant under $p4g$.
This wallpaper group contains the fourfold rotations $\{4^\pm|0\ 0\}$ and the twofold rotation $\{2|0\ 0\}$, as well as the two orthogonal glide operations $\{m_{10}|1/2\ 1/2\}$ and $\{m_{01}|1/2\ 1/2\}$.
The combination of these operations also gives the diagonal mirror operations $\{m_{11}|1/2\ 1/2\}$ and $\{m_{1\bar{1}}|1/2\ 1/2\}$.
As a result of the nonsymmorphic wallpaper-group symmetries, wallpaper fermions exhibit characteristic symmetry-enforced degeneracies, such as a fourfold degeneracy at the $\M$ point and twofold degeneracies along the $\X\M$ line.


\begin{figure}
    \centering

    \subfloat[]{
        \includegraphics[keepaspectratio,scale=0.65]{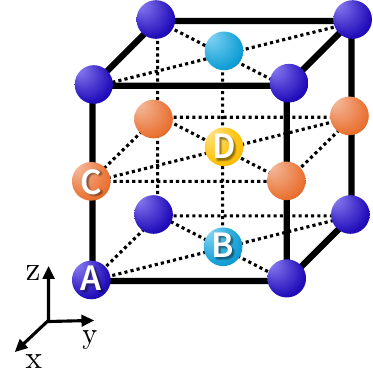}
        \label{fig:model}
    }\hfil
    \subfloat[]{
        \includegraphics[keepaspectratio,scale=0.35]{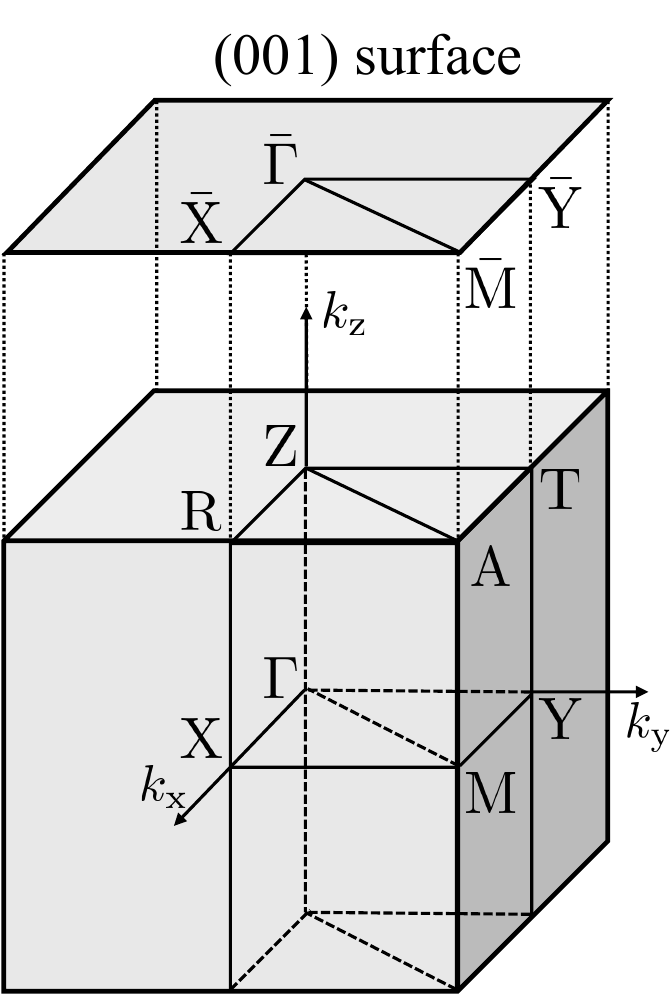}
        \label{fig:BZ}
    }

    \caption{(Color online)
        (a) Tetragonal four-sublattice model for wallpaper fermions~\cite{yoda2026double}.
        (b) Bulk Brillouin zone and the corresponding (001) surface Brillouin zones~\cite{Aroyo2011-cr, Aroyo2006-bi1, Aroyo2006-bi2, Aroyo2014brillouin}.
        In the system with the wallpaper group $p4g$, fourfold rotation symmetry makes the $\X$ and $\Y$ points symmetry-equivalent.
        When nematic superconductivity in the $\mathrm{E_g}$ or $\mathrm{E_u}$ representation is realized, this fourfold rotation symmetry can be spontaneously broken, so that these two points become inequivalent.
        The same applies to the corresponding momentum regions in the bulk Brillouin zone.
    }
    \label{fig:model_BZ}
\end{figure}

\subsection{Lattice model for wallpaper fermions} \label{sec:normal}

To describe the wallpaper fermions, we adopt the tetragonal four-sublattice model shown in Fig.~\ref{fig:model_BZ}\subref{fig:model}, originally introduced by B. J. Wieder \textit{et al}.~\cite{wieder2018wallpaper}.
The bulk crystalline symmetry corresponds to the space group $P4/mbm$ (No.~127), whose (001) surface preserves the wallpaper group $p4g$.
Figure~\ref{fig:model_BZ}\subref{fig:BZ} shows the bulk and surface Brillouin zones.
This model is described by three degrees of freedom: spin [$\up$ and $\down$], layer sublattice [$\mathrm{A(B)}$ and $\mathrm{C(D)}$], and in-plane sublattice [$\mathrm{A(C)}$ and $\mathrm{B(D)}$], which are represented by Pauli matrices $s_\nu$, $\rho_\nu$, and $\sigma_\nu$ ($\nu=0,\mathrm{x,\,y,\,z}$), respectively.
Throughout this work, the lattice constants are set to unity and we take the sublattice positions $\bm{\tau}_X$ ($X=\mathrm{A,B,C,D}$) as $\bm{\tau}_\mathrm{A}\coloneq(0,0,0)$, $\bm{\tau}_\mathrm{B}\coloneq(1/2,1/2,0)$, $\bm{\tau}_\mathrm{C}\coloneq(0,0,1/2)$, and $\bm{\tau}_\mathrm{D}\coloneq(1/2,1/2,1/2)$.

In this work, we employ the tight-binding Hamiltonian $H_\wp(\bm{k})$ introduced in Ref.~\onlinecite{yoda2026double} and give its explicit form below.
The in-plane part consists of the nearest-neighbor hopping
\begin{align}
    H^{(1)}_{\parallel}(\bm{k})\coloneq
     & t_1\cos{\frac{\kx}{2}}\cos{\frac{\ky}{2}}[s_0\rho_0\sigmax] \notag              \\
     & +v_{\mathrm{r}1}\cos{\frac{\kx}{2}}\cos{\frac{\ky}{2}}[\sz\rho_0\sigmay] \notag \\
     & +v_{\mathrm{s}1}\sin{\frac{\kx}{2}}\cos{\frac{\ky}{2}}[\sy\rhoz\sigmax] \notag  \\
     & -v_{\mathrm{s}1}\cos{\frac{\kx}{2}}\sin{\frac{\ky}{2}}[\sx\rhoz\sigmax]
    \label{eq:on_plane_1}
\end{align}
and the next-nearest-neighbor hopping
\begin{align}
    H^{(2)}_{\parallel}(\bm{k})\coloneq
     & t_2\qty(\cos{\kx}+\cos{\ky})[s_0\rho_0\sigma_0] \notag                             \\
     & +v_{\mathrm{s}2}\qty(\sin{\kx}[\sx\rhoz\sigmaz]+\sin{\ky}[\sy\rhoz\sigmaz]) \notag \\
     & +v'_{\mathrm{s}2}\qty(\sin{\kx}[\sy\rhoz\sigma_0]-\sin{\ky}[\sx\rhoz\sigma_0]).
    \label{eq:on_plane_2}
\end{align}
The out-of-plane part is composed of the nearest-neighbor hopping
\begin{equation}
    H^{(1)}_{\perp}(\bm{k})\coloneq u_1\cos{\frac{\kz}{2}}[s_0\rhox\sigma_0]+u_2\sin{\frac{\kz}{2}}[s_0\rhoy\sigma_0],
    \label{eq:out_plane_1}
\end{equation}
the next-nearest-neighbor hopping
\begin{align}
    H^{(2)}_{\perp}(\bm{k})\coloneq
     & w_1\cos{\frac{\kx}{2}}\cos{\frac{\ky}{2}}\cos{\frac{\kz}{2}}[s_0\rhox\sigmax] \notag  \\
     & +w_2\cos{\frac{\kx}{2}}\cos{\frac{\ky}{2}}\cos{\frac{\kz}{2}}[\sz\rhox\sigmay] \notag \\
     & +w_3\cos{\frac{\kx}{2}}\cos{\frac{\ky}{2}}\sin{\frac{\kz}{2}}[s_0\rhoy\sigmax] \notag \\
     & +w_4\cos{\frac{\kx}{2}}\cos{\frac{\ky}{2}}\sin{\frac{\kz}{2}}[\sz\rhoy\sigmay] \notag \\
     & +w_5\cos{\frac{\kx}{2}}\sin{\frac{\ky}{2}}\sin{\frac{\kz}{2}}[\sy\rhox\sigmay] \notag \\
     & +w_5\sin{\frac{\kx}{2}}\cos{\frac{\ky}{2}}\sin{\frac{\kz}{2}}[\sx\rhox\sigmay] \notag \\
     & +w_6\sin{\frac{\kx}{2}}\cos{\frac{\ky}{2}}\cos{\frac{\kz}{2}}[\sx\rhoy\sigmay] \notag \\
     & +w_6\cos{\frac{\kx}{2}}\sin{\frac{\ky}{2}}\cos{\frac{\kz}{2}}[\sy\rhoy\sigmay],
    \label{eq:out_plane_2}
\end{align}
the next-next-nearest-neighbor hopping
\begin{multline}
    H^{(3)}_{\perp}(\bm{k})\coloneq v_1(\cos{\kx}+\cos{\ky})\cos{\frac{\kz}{2}}[s_0\rhox\sigma_0] \\
    +v_2(\cos{\kx}+\cos{\ky})\sin{\frac{\kz}{2}}[s_0\rhoy\sigma_0],
    \label{eq:out_plane_3}
\end{multline}
and the intra-sublattice hopping
\begin{equation}
    H^{(4)}_{\perp}(\bm{k})\coloneq p_1\cos{\kz}[s_0\rho_0\sigma_0]+p_2\sin{\kz}[\sz\rhoz\sigmaz].
    \label{eq:out_plane_4}
\end{equation}
All hopping coefficients are taken to be real and hereafter, we take $t_1$ as the energy unit, i.e., $t_1=1$.
Here, we adopt the tensor-product convention $s_\mu\rho_\nu\sigma_\lambda\coloneq s_\mu\otimes\rho_\nu\otimes\sigma_\lambda$, corresponding to the basis $(c_{\bm{k},\mathrm{A},\up},\, c_{\bm{k},\mathrm{B},\up},\, c_{\bm{k},\mathrm{C},\up},\, c_{\bm{k},\mathrm{D},\up},\,\allowbreak c_{\bm{k},\mathrm{A},\down},\, c_{\bm{k},\mathrm{B},\down},\, c_{\bm{k},\mathrm{C},\down},\, c_{\bm{k},\mathrm{D},\down})$.
In the present model, the terms containing the spin-space identity matrix $s_0$ describe spin-independent hoppings, whereas those involving $\sx$, $\sy$, or $\sz$ describe spin--orbit terms.
Combining all contributions, the tight-binding Hamiltonian is given by
\begin{multline}
    H_\wp(\bm{k})\coloneq H^{(1)}_{\parallel}(\bm{k})+H^{(2)}_{\parallel}(\bm{k}) \\
    +H^{(1)}_{\perp}(\bm{k})+H^{(2)}_{\perp}(\bm{k})+H^{(3)}_{\perp}(\bm{k})+H^{(4)}_{\perp}(\bm{k}),
    \label{eq:wp}
\end{multline}
which is used as the normal Hamiltonian in the following analysis.


\begin{table*}
    \begin{ruledtabular}
        \centering
        \caption{
            The irreps and their character of the point group $\mathrm{D_{4h}}$, the classification of the pair potential, and the effective interaction for each basis pair potential.
        }
        \begin{tabular}{crrrcll}
            Irrep             & $2C_4$ & $2C'_2$ & $I$  & Basis                    & \multicolumn{1}{c}{Pair potential}                                                                                                                                         & \multicolumn{1}{c}{Effective interaction}                                         \\ \hline
            $\mathrm{A_{1g}}$ & $1$    & $1$     & $1$  & $\kx^2+\ky^2,\ \kz^2$    & $\hat{\Delta}^{(\mathrm{A_{1g}})}_1=s_0\rho_0\sigma_0$                                                                                                                     & $V^{(\mathrm{A_{1g}})}_1=\frac{1}{4}V_\mathrm{on}+V_\parallel+\frac{1}{2}V_\perp$ \\
                              &        &         &      &                          & $\hat{\Delta}^{(\mathrm{A_{1g}})}_2=\cos{\frac{\kx}{2}}\cos{\frac{\ky}{2}}s_0\rho_0\sigmax$                                                                                & $V^{(\mathrm{A_{1g}})}_2=V'_\parallel$                                            \\
                              &        &         &      &                          & $\hat{\Delta}^{(\mathrm{A_{1g}})}_3=\cos{\frac{\kx}{2}}\cos{\frac{\ky}{2}}\sz\rho_0\sigmay$                                                                                & $V^{(\mathrm{A_{1g}})}_3=V'_\parallel$                                            \\
                              &        &         &      &                          & $\hat{\Delta}^{(\mathrm{A_{1g}})}_4=\frac{1}{\sqrt{2}}\qty(\sin{\frac{\kx}{2}}\cos{\frac{\ky}{2}}\sy\rhoz\sigmax-\cos{\frac{\kx}{2}}\sin{\frac{\ky}{2}}\sx\rhoz\sigmax)$   & $V^{(\mathrm{A_{1g}})}_4=V'_\parallel$                                            \\
                              &        &         &      &                          & $\hat{\Delta}^{(\mathrm{A_{1g}})}_5=\cos{\frac{\kz}{2}}s_0\rhox\sigma_0$                                                                                                   & $V^{(\mathrm{A_{1g}})}_5=\frac{1}{2}V'_\perp$                                     \\
                              &        &         &      &                          & $\hat{\Delta}^{(\mathrm{A_{1g}})}_6=\sin{\frac{\kz}{2}}s_0\rhoy\sigma_0$                                                                                                   & $V^{(\mathrm{A_{1g}})}_6=\frac{1}{2}V'_\perp$                                     \\ [5pt]
            $\mathrm{A_{2g}}$ & $1$    & $-1$    & $1$  & $\kx\ky(\kx^2-\ky^2)$    & $\hat{\Delta}^{(\mathrm{A_{2g}})}_1=s_0\rho_0\sigmaz$                                                                                                                      & $V^{(\mathrm{A_{2g}})}_1=\frac{1}{4}V_\mathrm{on}-V_\parallel+\frac{1}{2}V_\perp$ \\
                              &        &         &      &                          & $\hat{\Delta}^{(\mathrm{A_{2g}})}_2=\frac{1}{\sqrt{2}}\qty(-\sin{\frac{\kx}{2}}\cos{\frac{\ky}{2}}\sx\rhoz\sigmax-\cos{\frac{\kx}{2}}\sin{\frac{\ky}{2}}\sy\rhoz\sigmax)$  & $V^{(\mathrm{A_{2g}})}_2=V'_\parallel$                                            \\
                              &        &         &      &                          & $\hat{\Delta}^{(\mathrm{A_{2g}})}_3=\cos{\frac{\kz}{2}}s_0\rhox\sigmaz$                                                                                                    & $V^{(\mathrm{A_{2g}})}_3=\frac{1}{2}V'_\perp$                                     \\
                              &        &         &      &                          & $\hat{\Delta}^{(\mathrm{A_{2g}})}_4=\sin{\frac{\kz}{2}}s_0\rhoy\sigmaz$                                                                                                    & $V^{(\mathrm{A_{2g}})}_4=\frac{1}{2}V'_\perp$                                     \\ [5pt]
            $\mathrm{B_{1g}}$ & $-1$   & $1$     & $1$  & $\kx^2-\ky^2$            & $\hat{\Delta}^{(\mathrm{B_{1g}})}_1=\frac{1}{\sqrt{2}}\qty(\sin{\frac{\kx}{2}}\cos{\frac{\ky}{2}}\sy\rhoz\sigmax+\cos{\frac{\kx}{2}}\sin{\frac{\ky}{2}}\sx\rhoz\sigmax)$   & $V^{(\mathrm{B_{1g}})}_1=V'_\parallel$                                            \\ [5pt]
            $\mathrm{B_{2g}}$ & $-1$   & $-1$    & $1$  & $\kx\ky$                 & $\hat{\Delta}^{(\mathrm{B_{2g}})}_1=\sin{\frac{\kx}{2}}\sin{\frac{\ky}{2}}s_0\rho_0\sigmax$                                                                                & $V^{(\mathrm{B_{2g}})}_1=V'_\parallel$                                            \\
                              &        &         &      &                          & $\hat{\Delta}^{(\mathrm{B_{2g}})}_2=\sin{\frac{\kx}{2}}\sin{\frac{\ky}{2}}\sz\rho_0\sigmay$                                                                                & $V^{(\mathrm{B_{2g}})}_2=V'_\parallel$                                            \\
                              &        &         &      &                          & $\hat{\Delta}^{(\mathrm{B_{2g}})}_3=\frac{1}{\sqrt{2}}\qty(-\sin{\frac{\kx}{2}}\cos{\frac{\ky}{2}}\sx\rhoz\sigmax+\cos{\frac{\kx}{2}}\sin{\frac{\ky}{2}}\sy\rhoz\sigmax)$  & $V^{(\mathrm{B_{2g}})}_3=V'_\parallel$                                            \\ [5pt]
            $\mathrm{E_g}$    & $0$    & $0$     & $1$  & $(\ky\kz,-\kx\kz)$       & $\hat{\Delta}^{(\mathrm{E_g})}_1=(\cos{\frac{\kx}{2}}\cos{\frac{\ky}{2}}\sy\rho_0\sigmay,-\cos{\frac{\kx}{2}}\cos{\frac{\ky}{2}}\sx\rho_0\sigmay)$                         & $V^{(\mathrm{E_g})}_1=V'_\parallel$                                               \\
                              &        &         &      &                          & $\hat{\Delta}^{(\mathrm{E_g})}_2=(-\sin{\frac{\kx}{2}}\sin{\frac{\ky}{2}}\sx\rho_0\sigmay,\sin{\frac{\kx}{2}}\sin{\frac{\ky}{2}}\sy\rho_0\sigmay)$                         & $V^{(\mathrm{E_g})}_2=V'_\parallel$                                               \\
                              &        &         &      &                          & $\hat{\Delta}^{(\mathrm{E_g})}_3=(\sin{\frac{\kx}{2}}\cos{\frac{\ky}{2}}s_0\rhoz\sigmay,\cos{\frac{\kx}{2}}\sin{\frac{\ky}{2}}s_0\rhoz\sigmay)$                            & $V^{(\mathrm{E_g})}_3=V'_\parallel$                                               \\
                              &        &         &      &                          & $\hat{\Delta}^{(\mathrm{E_g})}_4=(\sin{\frac{\kx}{2}}\cos{\frac{\ky}{2}}\sz\rhoz\sigmax,\cos{\frac{\kx}{2}}\sin{\frac{\ky}{2}}\sz\rhoz\sigmax)$                            & $V^{(\mathrm{E_g})}_4=V'_\parallel$                                               \\ [5pt]
            $\mathrm{A_{1u}}$ & $1$    & $1$     & $-1$ & $\kx\ky\kz(\kx^2-\ky^2)$ & $\hat{\Delta}^{(\mathrm{A_{1u}})}_1=s_0\rhoz\sigmaz$                                                                                                                       & $V^{(\mathrm{A_{1u}})}_1=\frac{1}{4}V_\mathrm{on}-V_\parallel-\frac{1}{2}V_\perp$ \\
                              &        &         &      &                          & $\hat{\Delta}^{(\mathrm{A_{1u}})}_2=\frac{1}{\sqrt{2}}\qty(\sin{\frac{\kx}{2}}\cos{\frac{\ky}{2}}\sx\rho_0\sigmax+\cos{\frac{\kx}{2}}\sin{\frac{\ky}{2}}\sy\rho_0\sigmax)$ & $V^{(\mathrm{A_{1u}})}_2=V'_\parallel$                                            \\
                              &        &         &      &                          & $\hat{\Delta}^{(\mathrm{A_{1u}})}_3=\cos{\frac{\kz}{2}}\sz\rhoy\sigma_0$                                                                                                   & $V^{(\mathrm{A_{1u}})}_3=\frac{1}{2}V'_\perp$                                     \\
                              &        &         &      &                          & $\hat{\Delta}^{(\mathrm{A_{1u}})}_4=\sin{\frac{\kz}{2}}\sz\rhox\sigma_0$                                                                                                   & $V^{(\mathrm{A_{1u}})}_4=\frac{1}{2}V'_\perp$                                     \\ [5pt]
            $\mathrm{A_{2u}}$ & $1$    & $-1$    & $-1$ & $\kz$                    & $\hat{\Delta}^{(\mathrm{A_{2u}})}_1=s_0\rhoz\sigma_0$                                                                                                                      & $V^{(\mathrm{A_{2u}})}_1=\frac{1}{4}V_\mathrm{on}+V_\parallel-\frac{1}{2}V_\perp$ \\
                              &        &         &      &                          & $\hat{\Delta}^{(\mathrm{A_{2u}})}_2=\cos{\frac{\kx}{2}}\cos{\frac{\ky}{2}}s_0\rhoz\sigmax$                                                                                 & $V^{(\mathrm{A_{2u}})}_2=V'_\parallel$                                            \\
                              &        &         &      &                          & $\hat{\Delta}^{(\mathrm{A_{2u}})}_3=\cos{\frac{\kx}{2}}\cos{\frac{\ky}{2}}\sz\rhoz\sigmay$                                                                                 & $V^{(\mathrm{A_{2u}})}_3=V'_\parallel$                                            \\
                              &        &         &      &                          & $\hat{\Delta}^{(\mathrm{A_{2u}})}_4=\frac{1}{\sqrt{2}}\qty(\sin{\frac{\kx}{2}}\cos{\frac{\ky}{2}}\sy\rho_0\sigmax-\cos{\frac{\kx}{2}}\sin{\frac{\ky}{2}}\sx\rho_0\sigmax)$ & $V^{(\mathrm{A_{2u}})}_4=V'_\parallel$                                            \\
                              &        &         &      &                          & $\hat{\Delta}^{(\mathrm{A_{2u}})}_5=\cos{\frac{\kz}{2}}\sz\rhoy\sigmaz$                                                                                                    & $V^{(\mathrm{A_{2u}})}_5=\frac{1}{2}V'_\perp$                                     \\
                              &        &         &      &                          & $\hat{\Delta}^{(\mathrm{A_{2u}})}_6=\sin{\frac{\kz}{2}}\sz\rhox\sigmaz$                                                                                                    & $V^{(\mathrm{A_{2u}})}_6=\frac{1}{2}V'_\perp$                                     \\ [5pt]
            $\mathrm{B_{1u}}$ & $-1$   & $1$     & $-1$ & $\kx\ky\kz$              & $\hat{\Delta}^{(\mathrm{B_{1u}})}_1=\sin{\frac{\kx}{2}}\sin{\frac{\ky}{2}}s_0\rhoz\sigmax$                                                                                 & $V^{(\mathrm{B_{1u}})}_1=V'_\parallel$                                            \\
                              &        &         &      &                          & $\hat{\Delta}^{(\mathrm{B_{1u}})}_2=\sin{\frac{\kx}{2}}\sin{\frac{\ky}{2}}\sz\rhoz\sigmay$                                                                                 & $V^{(\mathrm{B_{1u}})}_2=V'_\parallel$                                            \\
                              &        &         &      &                          & $\hat{\Delta}^{(\mathrm{B_{1u}})}_3=\frac{1}{\sqrt{2}}\qty(\sin{\frac{\kx}{2}}\cos{\frac{\ky}{2}}\sx\rho_0\sigmax-\cos{\frac{\kx}{2}}\sin{\frac{\ky}{2}}\sy\rho_0\sigmax)$ & $V^{(\mathrm{B_{1u}})}_3=V'_\parallel$                                            \\ [5pt]
            $\mathrm{B_{2u}}$ & $-1$   & $-1$    & $-1$ & $(\kx^2-\ky^2)\kz$       & $\hat{\Delta}^{(\mathrm{B_{2u}})}_1=\frac{1}{\sqrt{2}}\qty(\sin{\frac{\kx}{2}}\cos{\frac{\ky}{2}}\sy\rho_0\sigmax+\cos{\frac{\kx}{2}}\sin{\frac{\ky}{2}}\sx\rho_0\sigmax)$ & $V^{(\mathrm{B_{2u}})}_1=V'_\parallel$                                            \\ [5pt]
            $\mathrm{E_u}$    & $0$    & $0$     & $-1$ & $(\kx,\ky)$              & $\hat{\Delta}^{(\mathrm{E_u})}_1=(\cos{\frac{\kx}{2}}\cos{\frac{\ky}{2}}\sx\rhoz\sigmay,\cos{\frac{\kx}{2}}\cos{\frac{\ky}{2}}\sy\rhoz\sigmay)$                            & $V^{(\mathrm{E_u})}_1=V'_\parallel$                                               \\
                              &        &         &      &                          & $\hat{\Delta}^{(\mathrm{E_u})}_2=(\sin{\frac{\kx}{2}}\sin{\frac{\ky}{2}}\sy\rhoz\sigmay,\sin{\frac{\kx}{2}}\sin{\frac{\ky}{2}}\sx\rhoz\sigmay)$                            & $V^{(\mathrm{E_u})}_2=V'_\parallel$                                               \\
                              &        &         &      &                          & $\hat{\Delta}^{(\mathrm{E_u})}_3=(\cos{\frac{\kx}{2}}\sin{\frac{\ky}{2}}s_0\rho_0\sigmay,-\sin{\frac{\kx}{2}}\cos{\frac{\ky}{2}}s_0\rho_0\sigmay)$                         & $V^{(\mathrm{E_u})}_3=V'_\parallel$                                               \\
                              &        &         &      &                          & $\hat{\Delta}^{(\mathrm{E_u})}_4=(\cos{\frac{\kx}{2}}\sin{\frac{\ky}{2}}\sz\rho_0\sigmax,-\sin{\frac{\kx}{2}}\cos{\frac{\ky}{2}}\sz\rho_0\sigmax)$                         & $V^{(\mathrm{E_u})}_4=V'_\parallel$                                               \\
                              &        &         &      &                          & $\hat{\Delta}^{(\mathrm{E_u})}_5=(\cos{\frac{\kz}{2}}\sx\rhoy\sigmaz,\cos{\frac{\kz}{2}}\sy\rhoy\sigmaz)$                                                                  & $V^{(\mathrm{E_u})}_5=\frac{1}{2}V'_\perp$                                        \\
                              &        &         &      &                          & $\hat{\Delta}^{(\mathrm{E_u})}_6=(\cos{\frac{\kz}{2}}\sy\rhoy\sigma_0,-\cos{\frac{\kz}{2}}\sx\rhoy\sigma_0)$                                                               & $V^{(\mathrm{E_u})}_6=\frac{1}{2}V'_\perp$                                        \\
                              &        &         &      &                          & $\hat{\Delta}^{(\mathrm{E_u})}_7=(\sin{\frac{\kz}{2}}\sy\rhox\sigma_0,-\sin{\frac{\kz}{2}}\sx\rhox\sigma_0)$                                                               & $V^{(\mathrm{E_u})}_7=\frac{1}{2}V'_\perp$                                        \\
                              &        &         &      &                          & $\hat{\Delta}^{(\mathrm{E_u})}_8=(\sin{\frac{\kz}{2}}\sx\rhox\sigmaz,\sin{\frac{\kz}{2}}\sy\rhox\sigmaz)$                                                                  & $V^{(\mathrm{E_u})}_8=\frac{1}{2}V'_\perp$                                        \\
        \end{tabular}
        \label{tab:Delta}
    \end{ruledtabular}
\end{table*}

\subsection{Pairing interaction} \label{sec:SC}

For the electron--electron interaction, we include the pairings between in-plane nearest-neighbor sublattices [$\mathrm{A}(\mathrm{C})$ and $\mathrm{B}(\mathrm{D})$] and between out-of-plane nearest-neighbor sublattices [$\mathrm{A}(\mathrm{B})$ and $\mathrm{C}(\mathrm{D})$], as well as the on-site pairings discussed in Ref.~\onlinecite{yoda2026double}.
For a given sublattice $X$, the four in-plane nearest-neighbor sublattices $X'$ are specified by displacement vectors $\bm{d}^{(\mu)}_\parallel$ as
\begin{align}
     & \bm{d}^{(1)}_\parallel\coloneq\qty(\frac{1}{2},\frac{1}{2},0), &  & \bm{d}^{(2)}_\parallel\coloneq\qty(-\frac{1}{2},\frac{1}{2},0), \notag \\
     & \bm{d}^{(3)}_\parallel\coloneq-\bm{d}^{(1)}_\parallel,         &  & \bm{d}^{(4)}_\parallel\coloneq-\bm{d}^{(2)}_\parallel.
\end{align}
Similarly, the out-of-plane nearest-neighbor sublattices $X'$ are specified by the two displacement vectors $\bm{d}^{(\mu)}_\perp$ given by
\begin{equation}
    \bm{d}^{(1)}_\perp\coloneq\qty(0,0,\frac{1}{2}),\quad \bm{d}^{(2)}_\perp\coloneq-\bm{d}^{(1)}_\perp.
\end{equation}
In this work, we employ an interaction Hamiltonian that can be written in real space as
\begin{align}
    \mathcal{H}_\mathrm{int}\coloneq
      & V_\mathrm{on}\sum_{\bm{n},X}n^2_{\bm{n}+\bm{\tau}_X} \notag                                                                                                 \\
    + & 2V_\parallel \sum_{\bm{n},X}\sum_{\mu=1}^{4}c^\dag_{\bm{n}+\bm{\tau}_X,\up}c^\dag_{\bm{n}+\bm{\tau}_X,\down} \notag                                         \\
      & \qquad{}\times c_{\bm{n}+\bm{\tau}_X+\bm{d}^{(\mu)}_\parallel,\down}c_{\bm{n}+\bm{\tau}_X+\bm{d}^{(\mu)}_\parallel,\up} \notag                              \\
    + & 2V_\perp \sum_{\bm{n},X}\sum_{\mu=1}^{2}c^\dag_{\bm{n}+\bm{\tau}_X,\up}c^\dag_{\bm{n}+\bm{\tau}_X,\down} \notag                                             \\
      & \qquad{}\times c_{\bm{n}+\bm{\tau}_X+\bm{d}^{(\mu)}_\perp,\down}c_{\bm{n}+\bm{\tau}_X+\bm{d}^{(\mu)}_\perp,\up} \notag                                      \\
    + & 2V'_\parallel \sum_{\bm{n}}\sum_{X\in\{\mathrm{A},\mathrm{C}\}}\sum_{\mu=1}^{4}n_{\bm{n}+\bm{\tau}_X}n_{\bm{n}+\bm{\tau}_X+\bm{d}^{(\mu)}_\parallel} \notag \\
    + & 2V'_\perp \sum_{\bm{n}}\sum_{X\in\{\mathrm{A},\mathrm{B}\}}\sum_{\mu=1}^{2}n_{\bm{n}+\bm{\tau}_X}n_{\bm{n}+\bm{\tau}_X+\bm{d}^{(\mu)}_\perp},
    \label{eq:int_real}
\end{align}
where $\bm{n}$ is a lattice vector and $c^\dagger_{\bm{r},s}$ creates an electron with spin $s$ at position $\bm{r}$.
$n_{\bm{r}}\coloneq \sum_{s=\up,\down}n_{\bm{r},s}$ $(n_{\bm{r},s}\coloneq c^\dag_{\bm{r},s}c_{\bm{r},s})$ denotes the electron-density operator at position $\bm{r}$.
The first term describes an on-site density--density interaction.
We note that it contains a one-body term, $n_{\bm{r}}^2=n_{\bm{r}}+2n_{\bm{r},\up}n_{\bm{r},\down}$.
The resulting one-body term is absorbed into the chemical potential throughout this work.
The second and third terms are the in-plane and out-of-plane nearest-neighbor pair-hopping interactions between on-site Cooper pairs, respectively.
The last two terms contain off-site pairing interactions; the fourth (fifth) term represents in-plane (out-of-plane) nearest-neighbor density--density interactions.
For the density--density interactions, each nearest-neighbor bond is counted once with the summation conventions in Eq.~\eqref{eq:int_real}, and a negative (positive) interaction parameter corresponds to an attractive (repulsive) interaction.

Applying the Fourier transformation defined by
\begin{equation}
    c^\dag_{\bm{k},X,s}\coloneq\frac{1}{\sqrt{N}}\sum_{\bm{n}}e^{i\bm{k}\cdot(\bm{n}+\bm{\tau}_X)}c^\dag_{\bm{n}+\bm{\tau}_X,s},
    \label{eq:fourier}
\end{equation}
where $N$ is the number of unit cells and assuming Cooper pairs with no center-of-mass momentum, the interaction Hamiltonian in Eq.~\eqref{eq:int_real} can be written in momentum space as follows:
\begin{multline}
    \mathcal{H}_\mathrm{int}=\frac{1}{2N}\sum_{\bm{k},\bm{k}'}\sum_{\substack{\alpha_1,\beta_1 \\ \alpha_2,\beta_2}}V_{\alpha_1\beta_1,\alpha_2\beta_2}(\bm{k},\bm{k}') \\
    \times c^\dag_{\bm{k},\alpha_1}c^\dag_{-\bm{k},\beta_1}c_{-\bm{k}',\beta_2}c_{\bm{k}',\alpha_2},
    \label{eq:int_k}
\end{multline}
where $\alpha_j$ and $\beta_j$ collectively denote spin and sublattice indices.
In the following mean-field approximation, we retain only the superconducting pairing channel and neglect the Hartree--Fock contributions to the normal Hamiltonian.
The resulting BdG Hamiltonian is given by
\begin{equation}
    H_\BdG(\bm{k})\coloneq[H_\wp(\bm{k})-\mu]\tauz+\hat{\Delta}(\bm{k})\taux,
    \label{eq:BdG}
\end{equation}
where $\mu$ is the chemical potential and $\tau_\nu\ (\nu=0,\mathrm{x,\,y,\,z})$ are Pauli matrices in Nambu (particle-hole) space.
$\hat{\Delta}(\bm{k})$ denotes the pair potential, whose phase is chosen such that $\Theta\hat{\Delta}(\bm{k})\Theta^{-1}=\hat{\Delta}(-\bm{k})$, where $\Theta\coloneq-i\sy\mathcal{K}$ is a time-reversal operator, with $\mathcal{K}$ complex conjugation.
Here, the Nambu basis is taken as $(c_{\bm{k},X,\up},c_{\bm{k},X,\down},c_{-\bm{k},X,\down}^\dag,-c_{-\bm{k},X,\up}^\dag)$.

The pair potential $\hat{\Delta}(\bm{k})$ is expanded in terms of irreducible representations (irreps) of the point group $\mathrm{D_{4h}}$, which is the rotational part of the space group $P4/mbm$, as~\cite{sigrist1991phenomenological}
\begin{equation}
    \hat{\Delta}(\bm{k})=\sum_{\Gamma,\gamma}\sum_{i}{}^\gamma\!\eta^{(\Gamma)}_i{}^\gamma\!\hat{\Delta}^{(\Gamma)}_i(\bm{k}),
    \label{eq:expand_Delta}
\end{equation}
where ${}^\gamma\!\hat{\Delta}^{(\Gamma)}_i(\bm{k})$ is the $i$th pair potential belonging to irrep $\Gamma$.
The superscript $\gamma$ labels the different components of each basis in multidimensional irreps and is omitted for one-dimensional irreps.
${}^\gamma\!\eta^{(\Gamma)}_i$ denotes the momentum-independent expansion coefficient corresponding to ${}^\gamma\!\hat{\Delta}^{(\Gamma)}_i(\bm{k})$.
The pair potentials satisfy $\sy[{}^\gamma\!\hat{\Delta}^{(\Gamma)}_i(-\bm{k})]^\mathrm{t}\sy={}^\gamma\!\hat{\Delta}^{(\Gamma)}_i(\bm{k})$ as required by Fermi--Dirac statistics.
In the present system, restricting on-site, in-plane nearest-neighbor, and out-of-plane nearest-neighbor pairings, the allowed pair potentials are classified as listed in Table~\ref{tab:Delta}.

\section{Phase Diagram} \label{sec:phase_diagram}

In this section, we determine the superconducting phase diagram by solving the linearized gap equation for each irrep.
This analysis reveals the interaction parameter regimes in which the $\mathrm{A_{1u}}$ channel, which gives rise to the twisted surface spectrum~\cite{yoda2026double}, is stabilized.
We also identify the leading off-site pairing channels that become dominant in competition with the on-site pairing channels.

\subsection{Linearized gap equation for each irreducible representation} \label{sec:gap_eq}

First, we briefly summarize the linearized gap equation for each irrep, following Refs.~\onlinecite{sigrist1991phenomenological, nomoto2016classification, kawakami2018topological}.
Using the basis pair potentials ${}^\gamma\!\hat{\Delta}^{(\Gamma)}_i(\bm{k})$ introduced in Eq.~\eqref{eq:expand_Delta}, the pairing interaction $V_{\alpha_1\beta_1,\alpha_2\beta_2}(\bm{k},\bm{k}')$ in Eq.~\eqref{eq:int_k} can be expanded as
\begin{multline}
    V_{\alpha_1\beta_1,\alpha_2\beta_2}(\bm{k},\bm{k}')= \\
    \sum_{\Gamma,\gamma}\sum_{i}V_i^{(\Gamma)}[{}^\gamma\!\hat{\Delta}^{(\Gamma)}_i(\bm{k})i\sy]_{\alpha_1\beta_1}[{}^\gamma\!\hat{\Delta}^{(\Gamma)}_i(\bm{k}')i\sy]^\ast_{\alpha_2\beta_2},
\end{multline}
where $V_i^{(\Gamma)}$ can be regarded as the effective pairing interaction associated with the $i$th pair potential in the irrep $\Gamma$.
Near the transition temperature $\Tc$, the gap equation can be linearized and decomposed into each irrep $\Gamma$.
It is expressed as the eigenvalue equation
\begin{equation}
    M^{(\Gamma)}{}^\gamma\!\bm{\eta}^{(\Gamma)}=\lambda^{(\Gamma)}(T){}^\gamma\!\bm{\eta}^{(\Gamma)},
    \label{eq:gap_eq}
\end{equation}
where the superconducting phase transition in the channel $\Gamma$ occurs when the largest eigenvalue $\lambda^{(\Gamma)}_\mathrm{max}(T)\coloneq\max\{\lambda^{(\Gamma)}(T)\}$ satisfies $\lambda^{(\Gamma)}_\mathrm{max}(\Tc)=1$.
Here, we define the vector ${}^\gamma\!\bm{\eta}^{(\Gamma)}$ such that $[{}^\gamma\!\bm{\eta}^{(\Gamma)}]_i={}^\gamma\!\eta^{(\Gamma)}_i$.
The matrix $M^{(\Gamma)}$ is given by
\begin{multline}
    [M^{(\Gamma)}]_{ij}\coloneq -\frac{V_i^{(\Gamma)}}{N}\sum_{\bm{k}}\sum_{\ell,\ell'}\frac{1-n_\mathrm{F}(\xi_{\ell}(\bm{k}))-n_\mathrm{F}(\xi_{\ell'}(\bm{k}))}{\xi_{\ell}(\bm{k})+\xi_{\ell'}(\bm{k})} \\
    \times \Tr[[{}^\gamma\!\hat{\Delta}^{(\Gamma)}_i(\bm{k})]^\dag \P_\ell(\bm{k}) {}^\gamma\!\hat{\Delta}^{(\Gamma)}_j(\bm{k}) \P_{\ell'}(\bm{k})],
    \label{eq:gap_eq_matrix}
\end{multline}
where $\xi_\ell(\bm{k})$ is the $\ell$th eigenenergy measured from the chemical potential $\mu$ in the normal state, $n_\mathrm{F}(\xi)\coloneq(1+e^{\xi/\kb T})^{-1}$ is the Fermi--Dirac distribution function, and $\P_\ell(\bm{k})$ is the projection operator onto the $\ell$th eigenstate.
The leading pairing channel is identified as the irrep whose largest eigenvalue $\lambda^{(\Gamma)}_\mathrm{max}(T)$ first reaches unity as the temperature is lowered.


\begin{table}
    \begin{ruledtabular}
        \centering
        \caption{
            Parameter set for the normal states.
            All parameters are set to the values used in Ref.~\onlinecite{yoda2026double}.
        }
        \begin{tabular}{cccccccccc}
            $t_1$  & $v_{\mathrm{r}1}$ & $v_{\mathrm{s}1}$ & $t_2$   & $v_{\mathrm{s}2}$ & $v'_{\mathrm{s}2}$ & $u_1$ & $u_2$   & $w_1$   & $w_2$  \\ \hline
            $1.0$  & $0.25$            & $0.2$             & $0.5$   & $-0.2$            & $0.15$             & $0.3$ & $-0.45$ & $0.2$   & $0.05$ \\ [5pt]
            $w_3$  & $w_4$             & $w_5$             & $w_6$   & $v_1$             & $v_2$              & $p_1$ & $p_2$   & $\mu$   &        \\ \hline
            $0.05$ & $-0.02$           & $0.11$            & $-0.08$ & $0.2$             & $0.1$              & $0.0$ & $0.3$   & $-0.75$ &        \\
        \end{tabular}
        \label{tab:parameter}
    \end{ruledtabular}
\end{table}
\begin{figure*}
    \centering

    \subfloat[$\alpha_\parallel=0.0,\ \alpha_\perp=1.0.$]{
        \includegraphics[keepaspectratio,scale=0.35]{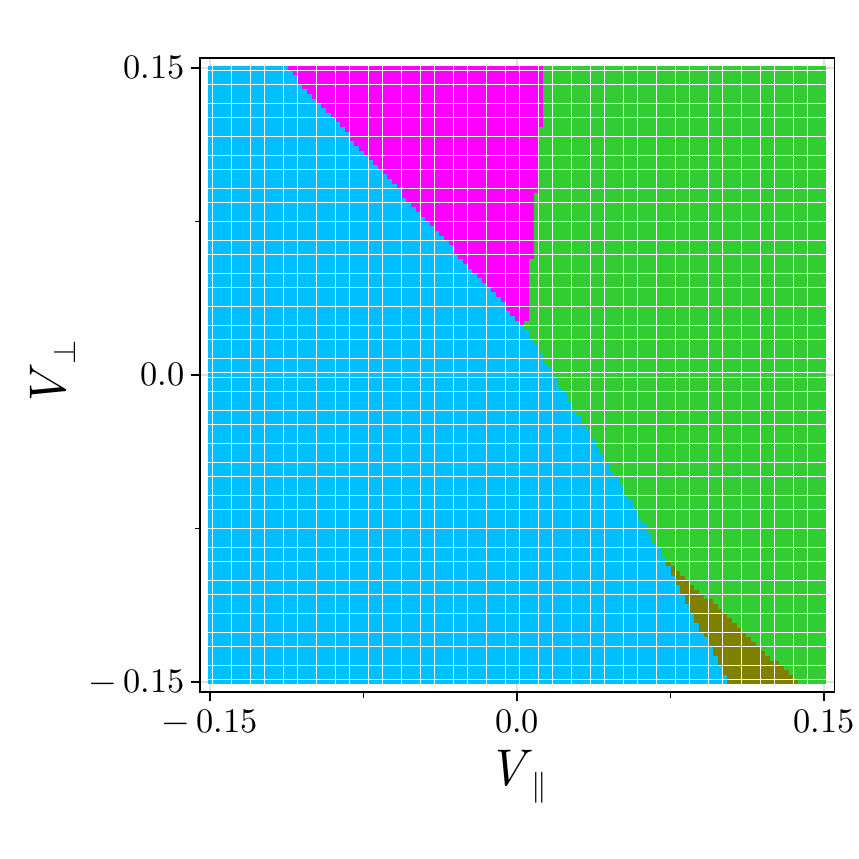}
        \label{fig:attractive_0.0_1.0}
    }\hfil
    \subfloat[$\alpha_\parallel=0.5,\ \alpha_\perp=1.0.$]{
        \includegraphics[keepaspectratio,scale=0.35]{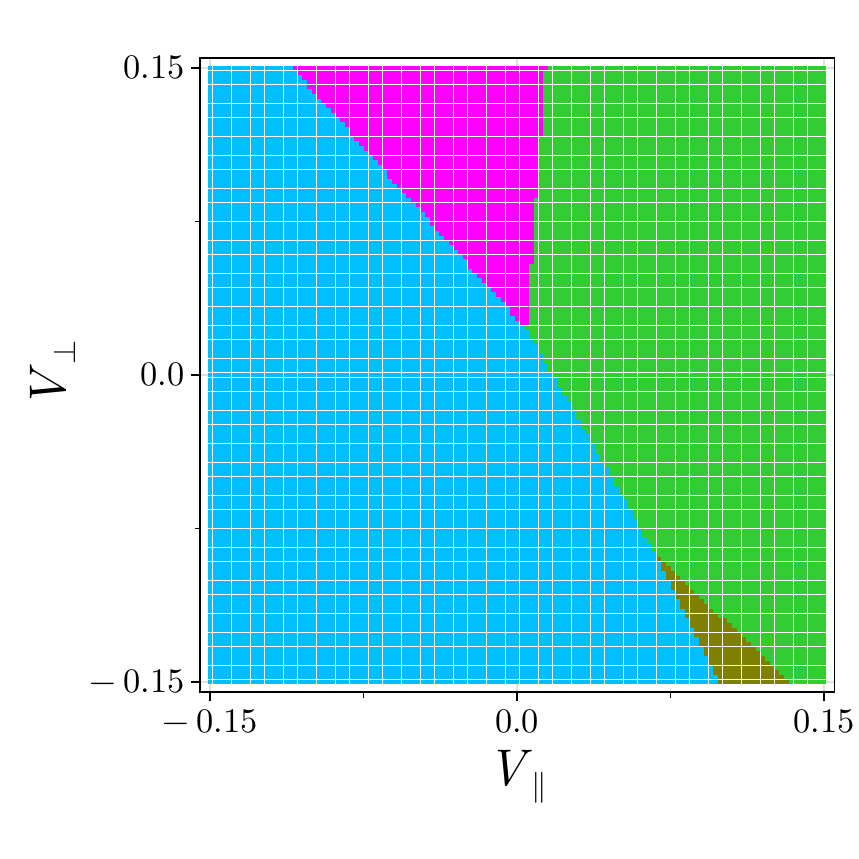}
        \label{fig:attractive_0.5_1.0}
    }\hfil
    \subfloat[$\alpha_\parallel=1.0,\ \alpha_\perp=1.0.$]{
        \includegraphics[keepaspectratio,scale=0.35]{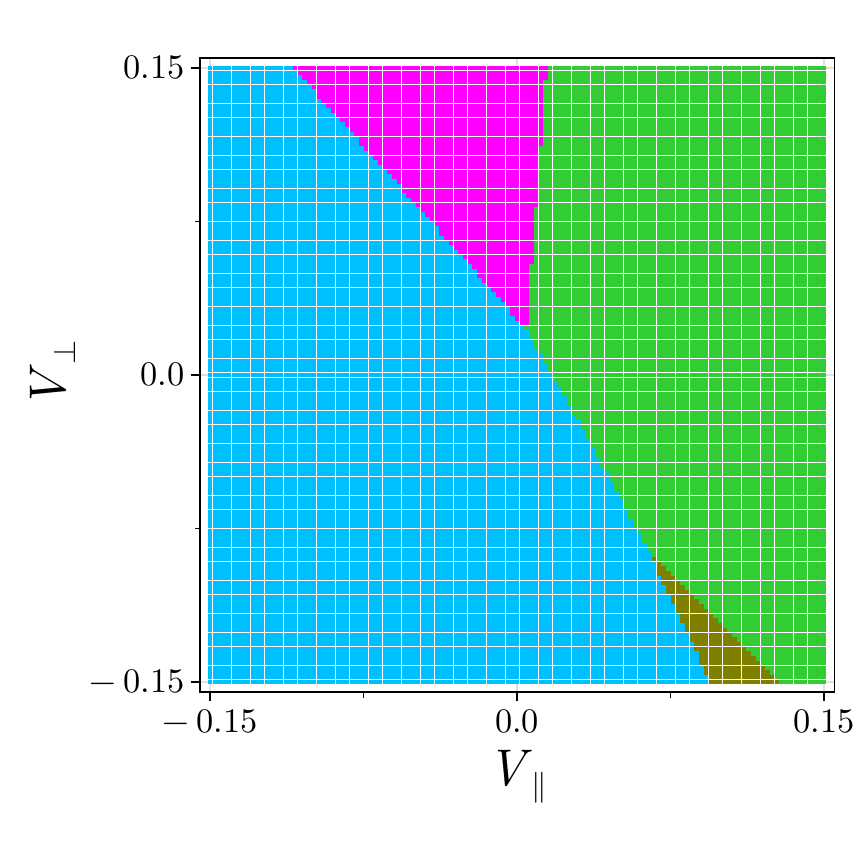}
        \label{fig:attractive_1.0_1.0}
    }\\
    \subfloat[$\alpha_\parallel=0.0,\ \alpha_\perp=0.5.$]{
        \includegraphics[keepaspectratio,scale=0.35]{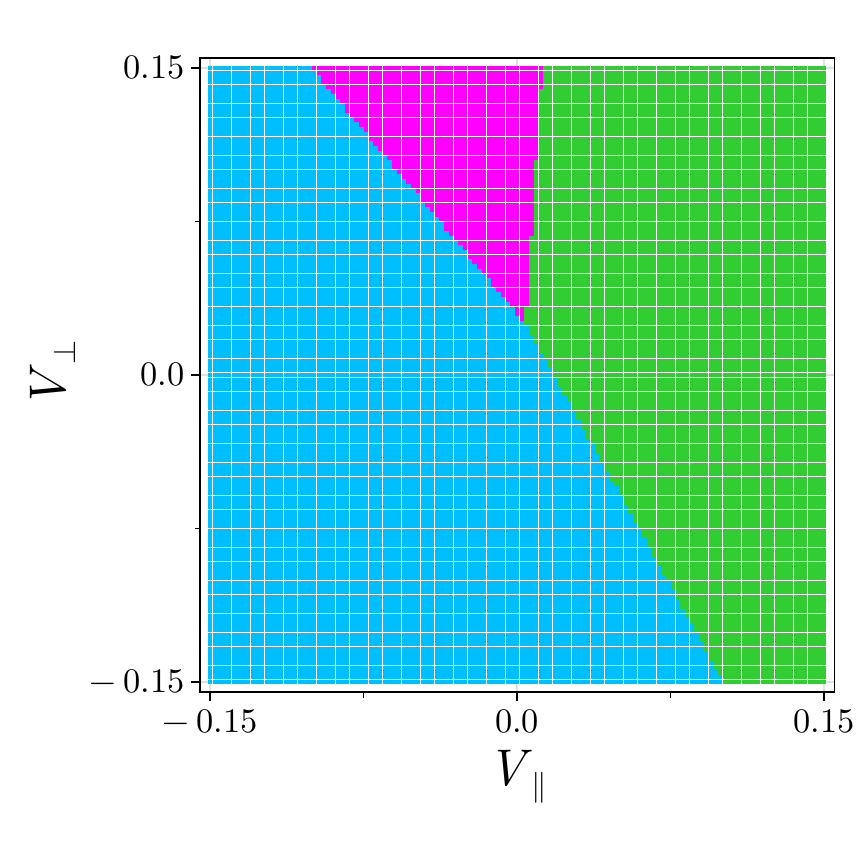}
        \label{fig:attractive_0.0_0.5}
    }\hfil
    \subfloat[$\alpha_\parallel=0.5,\ \alpha_\perp=0.5.$]{
        \includegraphics[keepaspectratio,scale=0.35]{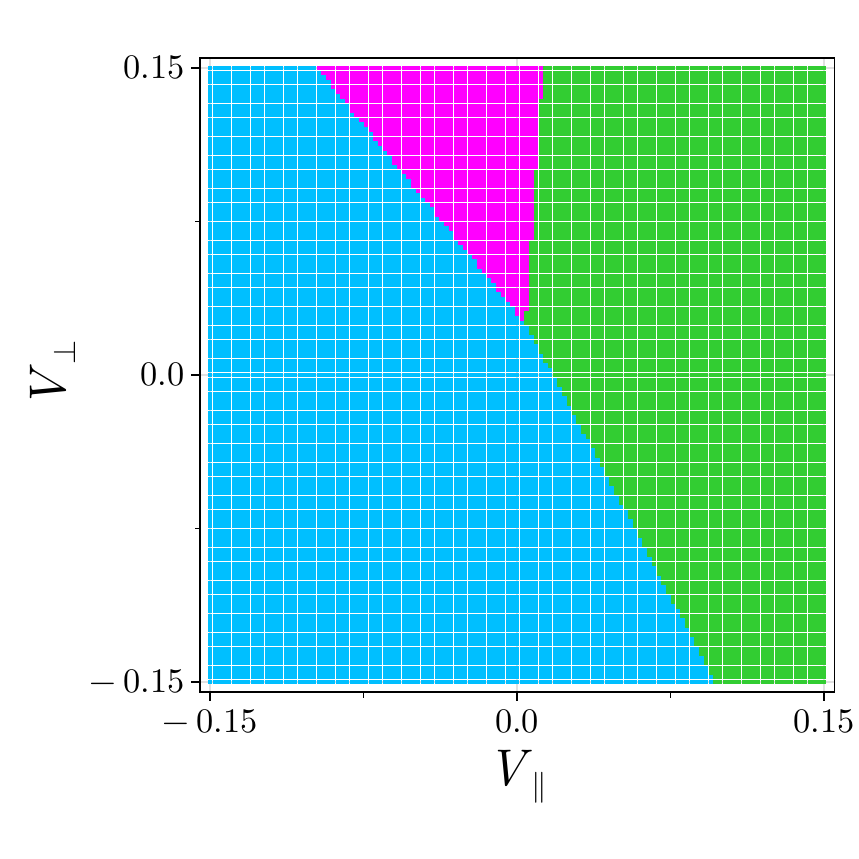}
        \label{fig:attractive_0.5_0.5}
    }\hfil
    \subfloat[$\alpha_\parallel=1.0,\ \alpha_\perp=0.5.$]{
        \includegraphics[keepaspectratio,scale=0.35]{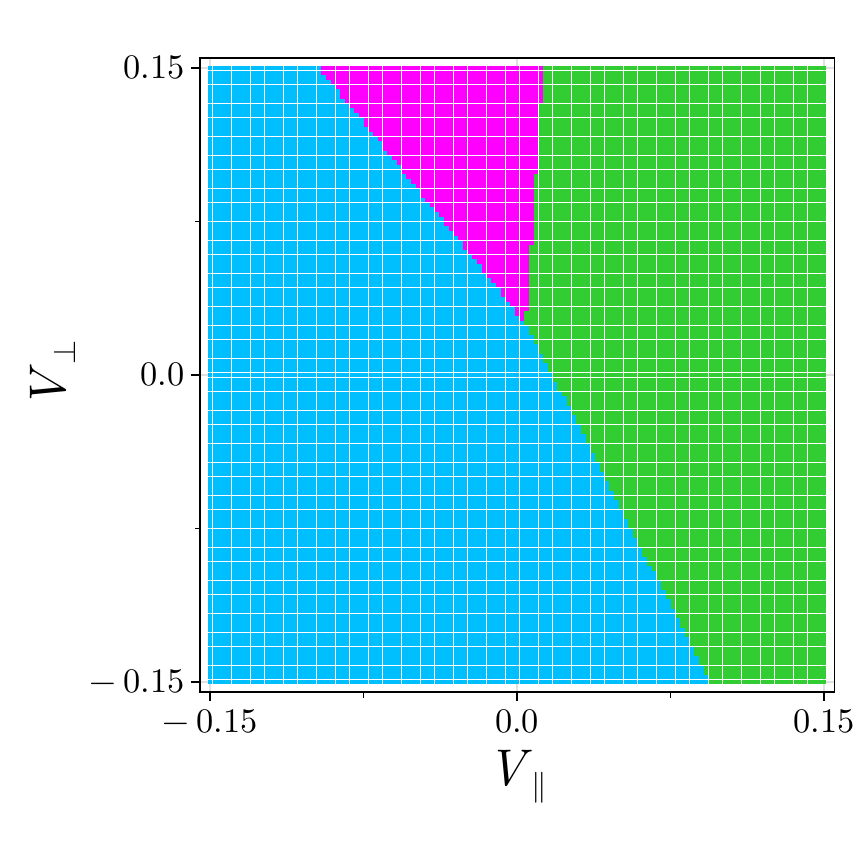}
        \label{fig:attractive_1.0_0.5}
    }\\
    \subfloat[$\alpha_\parallel=0.0,\ \alpha_\perp=0.0.$]{
        \includegraphics[keepaspectratio,scale=0.35]{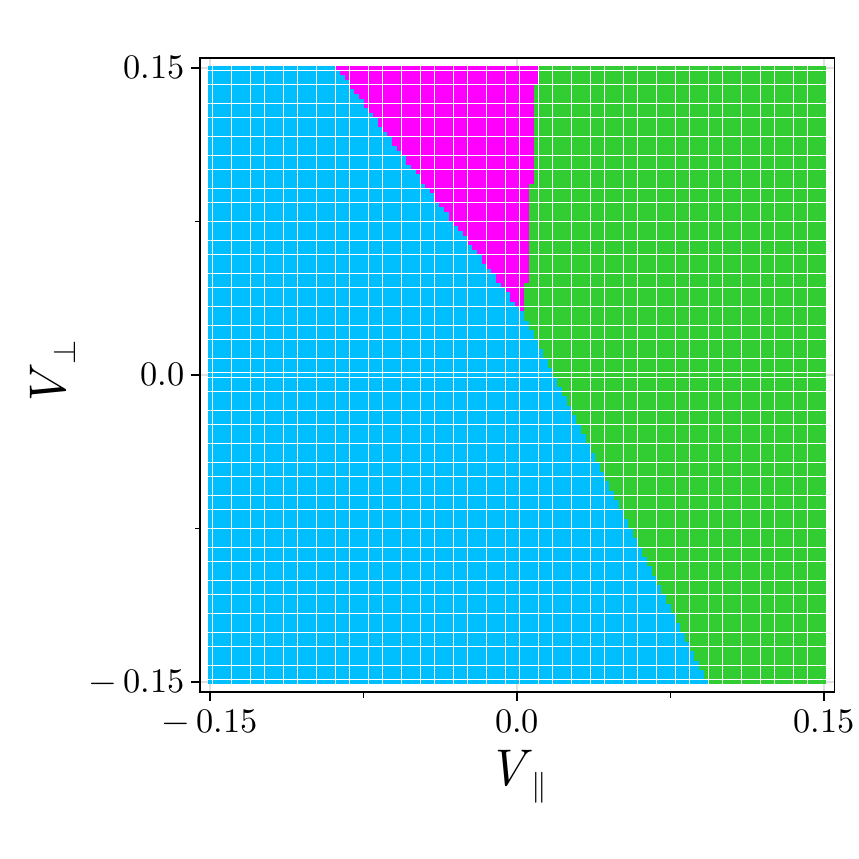}
        \label{fig:attractive_0.0_0.0}
    }\hfil
    \subfloat[$\alpha_\parallel=0.5,\ \alpha_\perp=0.0.$]{
        \includegraphics[keepaspectratio,scale=0.35]{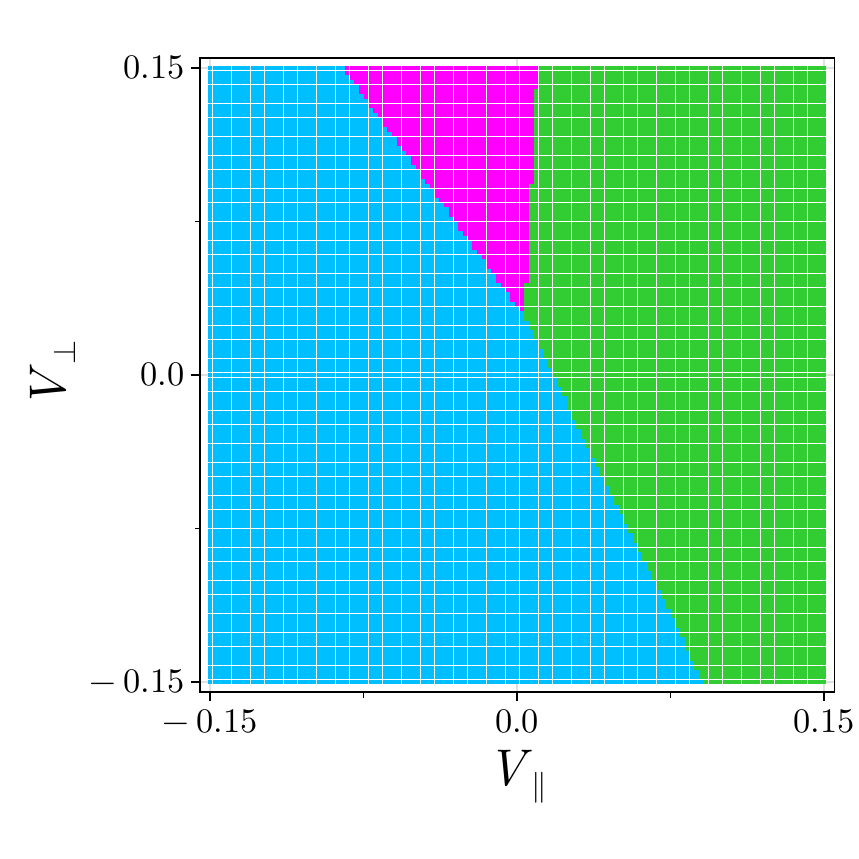}
        \label{fig:attractive_0.5_0.0}
    }\hfil
    \subfloat[$\alpha_\parallel=1.0,\ \alpha_\perp=0.0.$]{
        \includegraphics[keepaspectratio,scale=0.35]{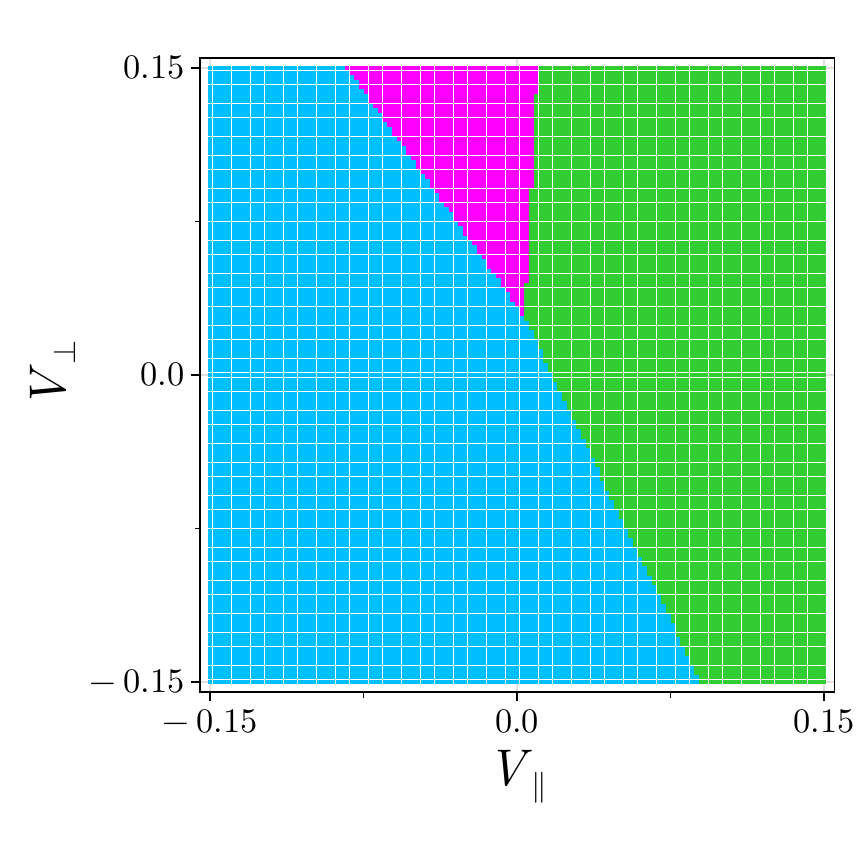}
        \label{fig:attractive_1.0_0.0}
    }\\
    \subfloat[Legend.]{
        \includegraphics[keepaspectratio,scale=0.8]{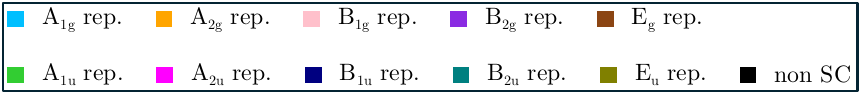}
        \label{fig:attractive_legend}
    }

    \caption{(Color online)
        Phase diagram for $V_\mathrm{on}=-0.1$.
    }
    \label{fig:phase_attractive}
\end{figure*}
\begin{figure*}
    \centering

    \subfloat[$\alpha_\parallel=0.0,\ \alpha_\perp=1.0.$]{
        \includegraphics[keepaspectratio,scale=0.35]{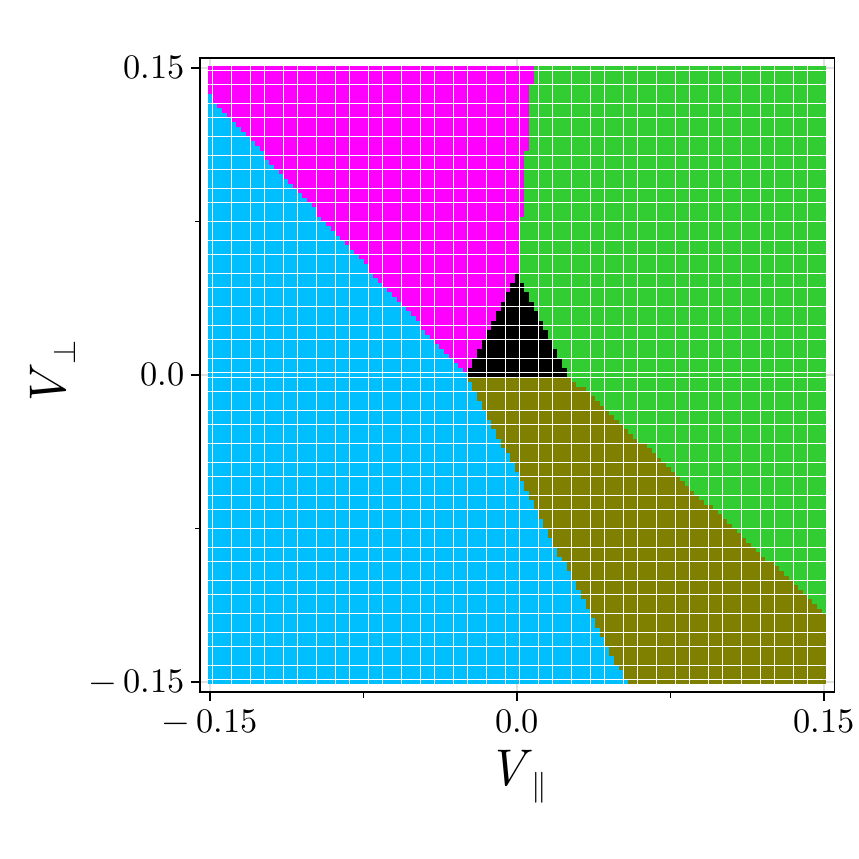}
        \label{fig:repulsive_0.0_1.0}
    }\hfil
    \subfloat[$\alpha_\parallel=0.5,\ \alpha_\perp=1.0.$]{
        \includegraphics[keepaspectratio,scale=0.35]{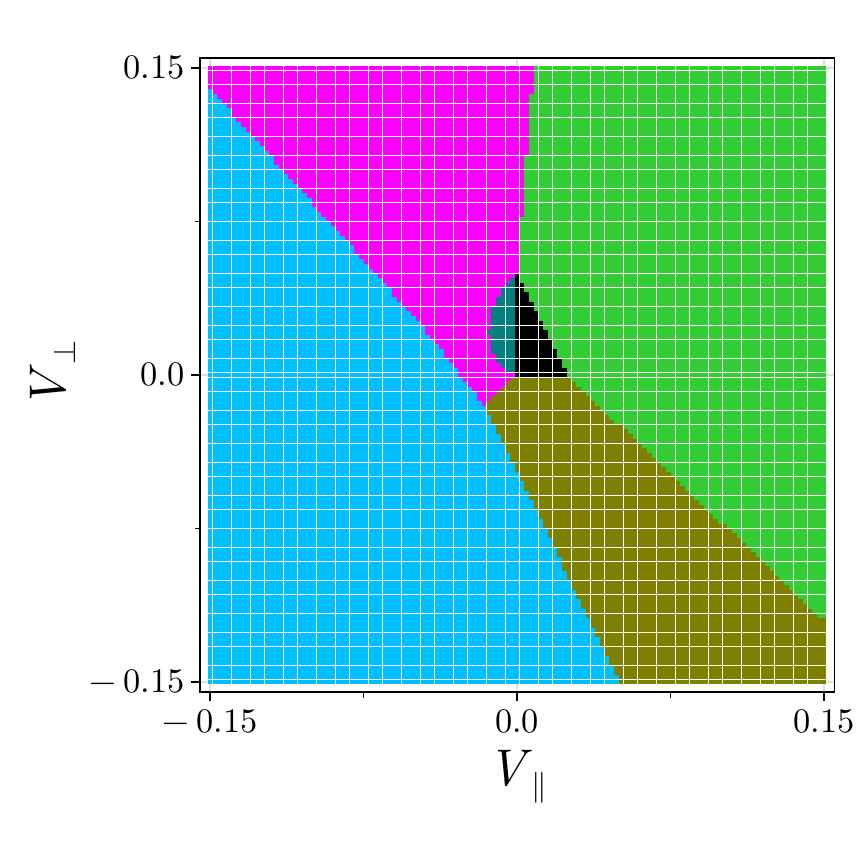}
        \label{fig:repulsive_0.5_1.0}
    }\hfil
    \subfloat[$\alpha_\parallel=1.0,\ \alpha_\perp=1.0.$]{
        \includegraphics[keepaspectratio,scale=0.35]{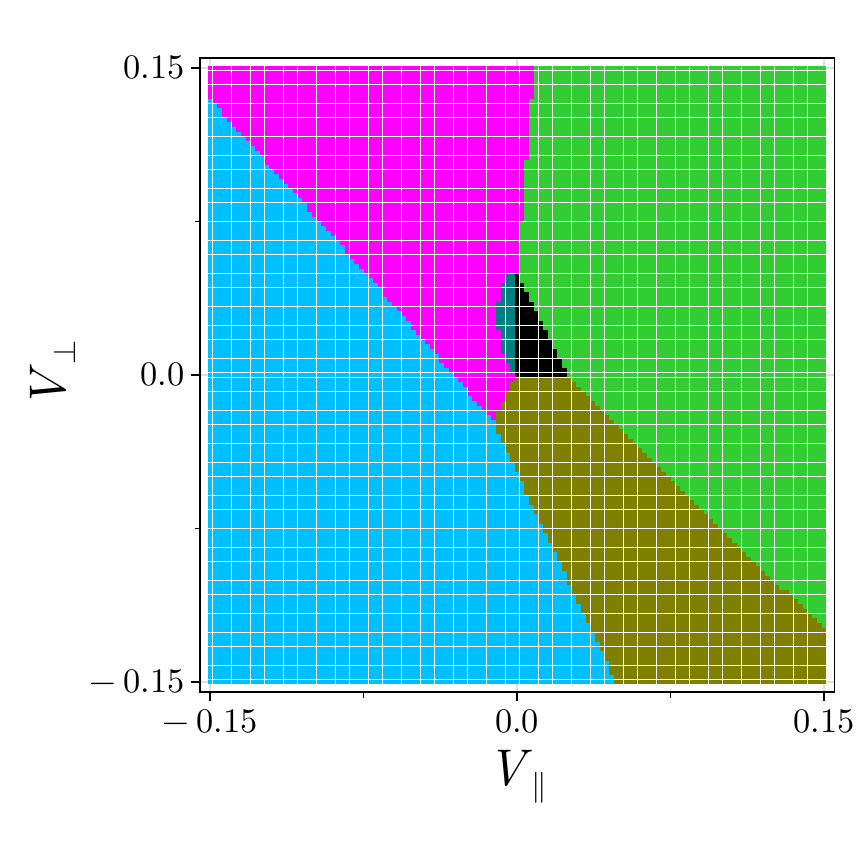}
        \label{fig:repulsive_1.0_1.0}
    }\\
    \subfloat[$\alpha_\parallel=0.0,\ \alpha_\perp=0.5.$]{
        \includegraphics[keepaspectratio,scale=0.35]{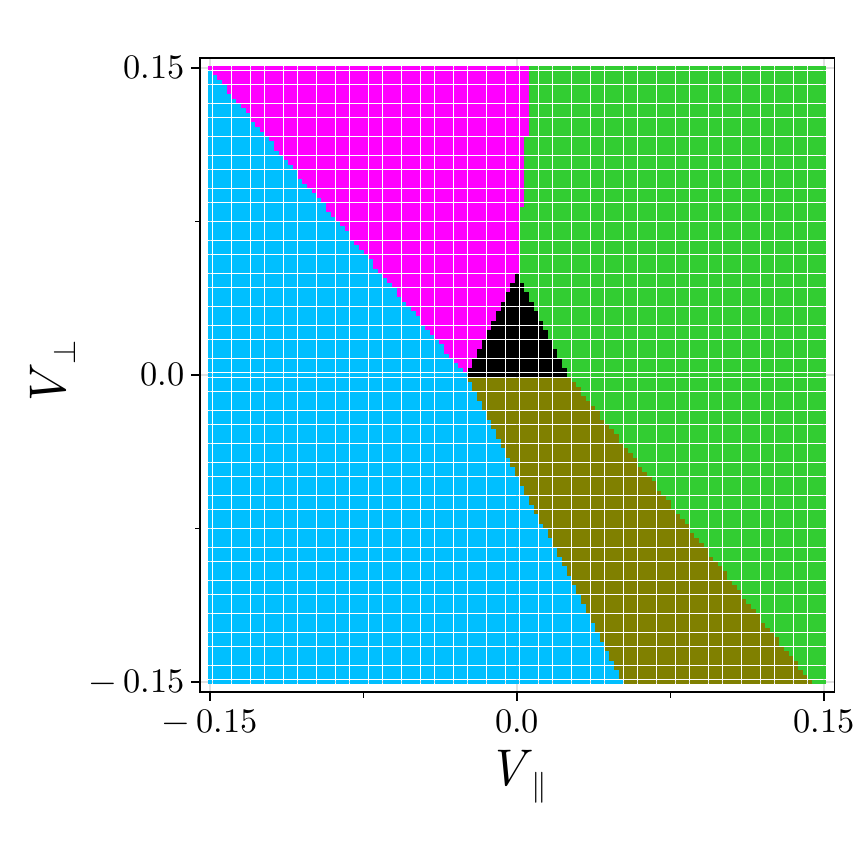}
        \label{fig:repulsive_0.0_0.5}
    }\hfil
    \subfloat[$\alpha_\parallel=0.5,\ \alpha_\perp=0.5.$]{
        \includegraphics[keepaspectratio,scale=0.35]{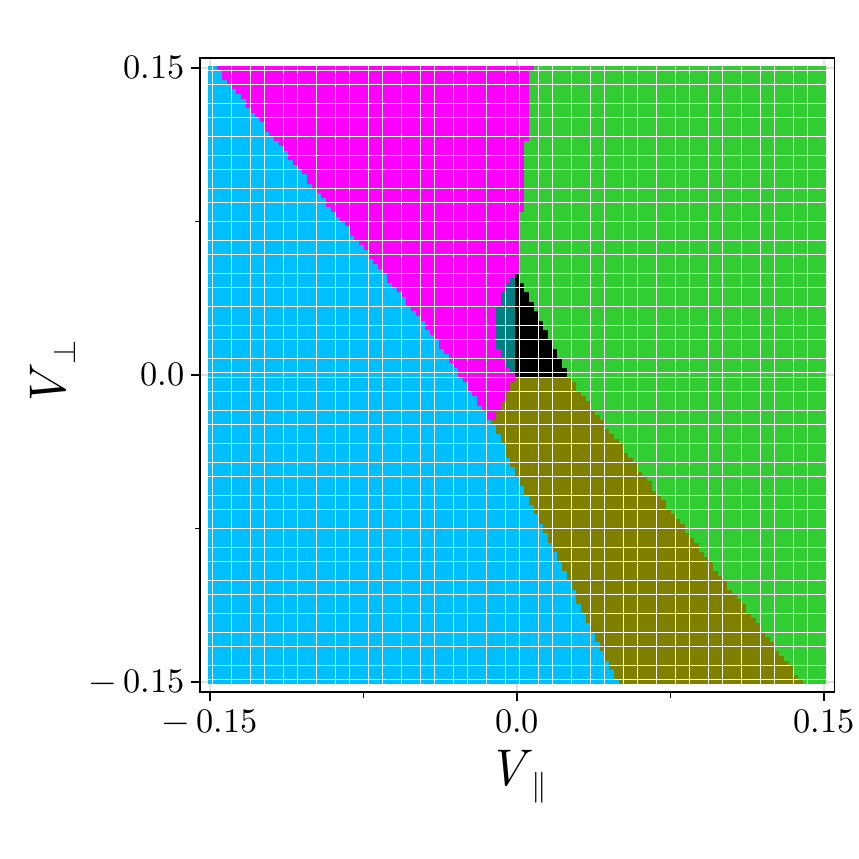}
        \label{fig:repulsive_0.5_0.5}
    }\hfil
    \subfloat[$\alpha_\parallel=1.0,\ \alpha_\perp=0.5.$]{
        \includegraphics[keepaspectratio,scale=0.35]{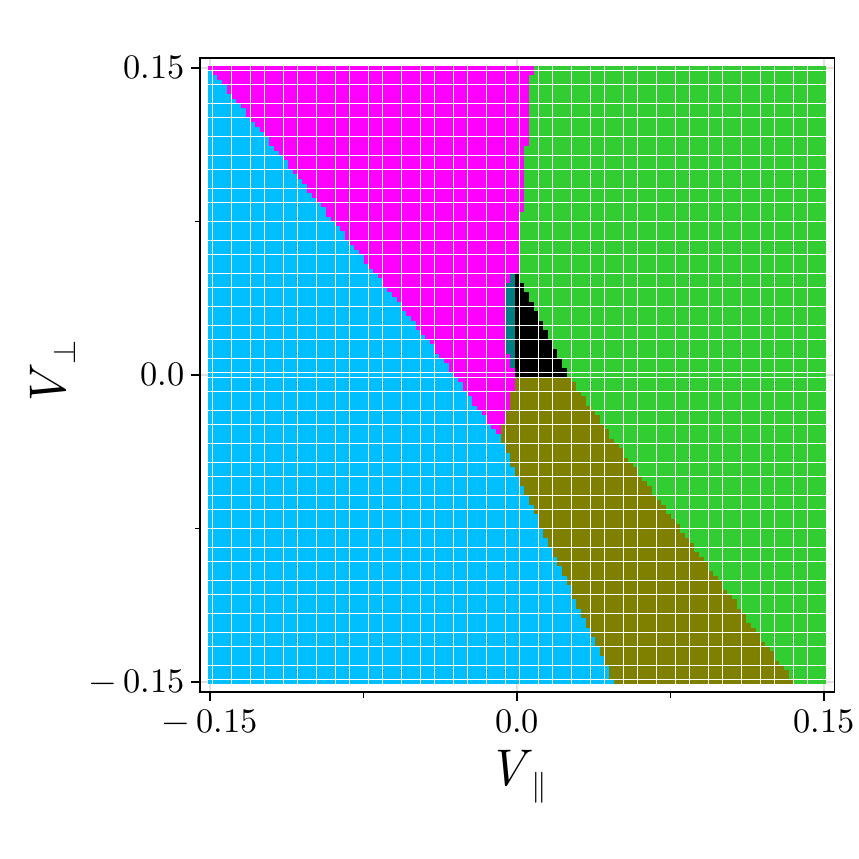}
        \label{fig:repulsive_1.0_0.5}
    }\\
    \subfloat[$\alpha_\parallel=0.0,\ \alpha_\perp=0.0.$]{
        \includegraphics[keepaspectratio,scale=0.35]{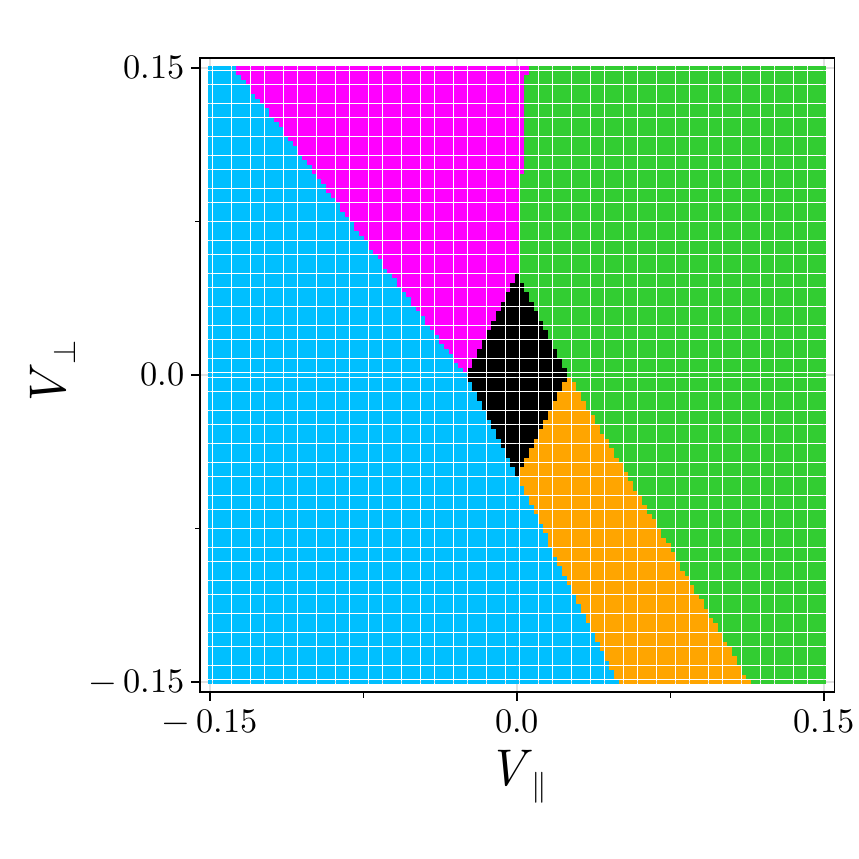}
        \label{fig:repulsive_0.0_0.0}
    }\hfil
    \subfloat[$\alpha_\parallel=0.5,\ \alpha_\perp=0.0.$]{
        \includegraphics[keepaspectratio,scale=0.35]{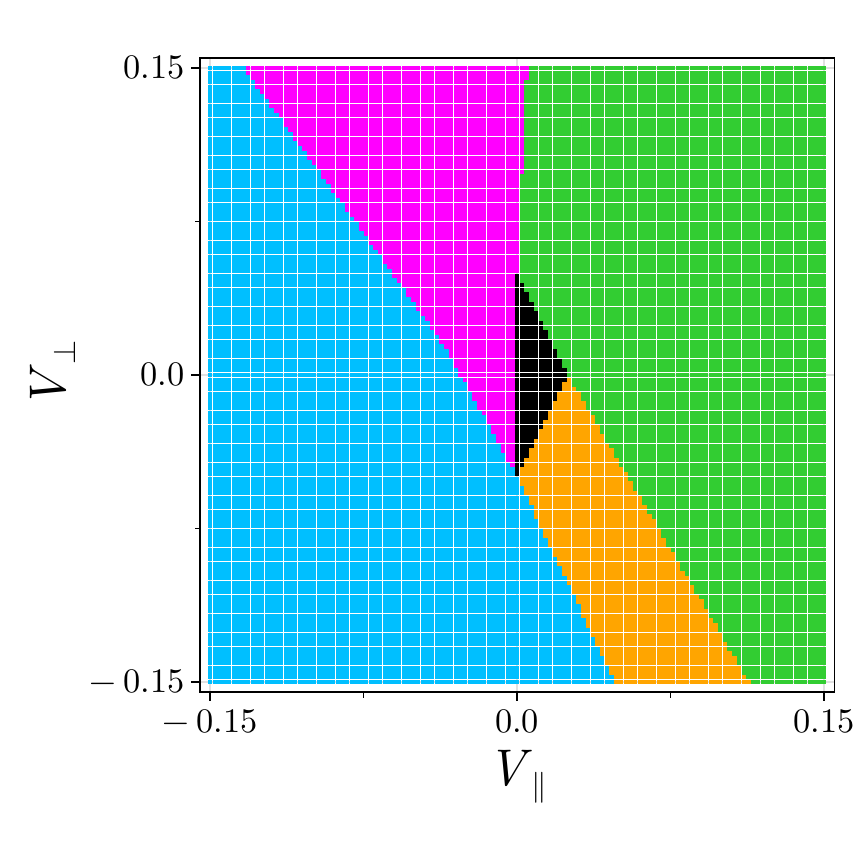}
        \label{fig:repulsive_0.5_0.0}
    }\hfil
    \subfloat[$\alpha_\parallel=1.0,\ \alpha_\perp=0.0.$]{
        \includegraphics[keepaspectratio,scale=0.35]{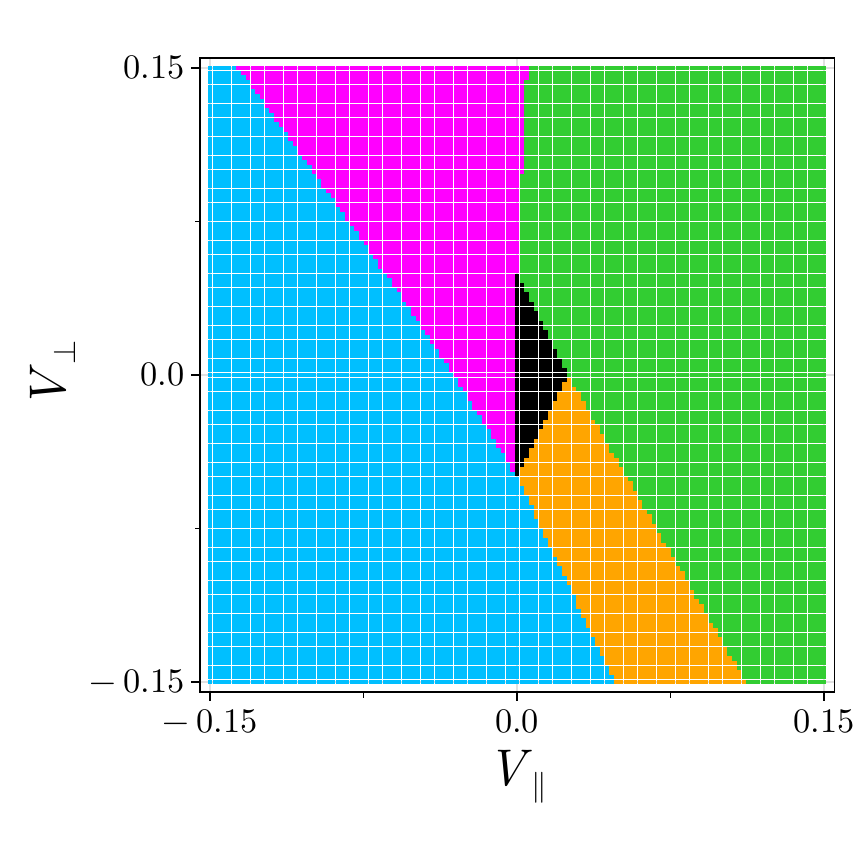}
        \label{fig:repulsive_1.0_0.0}
    }\\
    \subfloat[Legend.]{
        \includegraphics[keepaspectratio,scale=0.8]{./figure/legend.pdf}
        \label{fig:repulsive_legend}
    }

    \caption{(Color online)
        Phase diagram for $V_\mathrm{on}=0.1$.
    }
    \label{fig:phase_repulsive}
\end{figure*}

\subsection{Numerical results of the phase diagrams} \label{sec:phase_result}

In this subsection, we determine the pairing phase diagrams based on the above formulation.
For the present electron--electron interaction Hamiltonian in Eq.~\eqref{eq:int_real}, the effective pairing interactions $V_i^{(\Gamma)}$ are obtained as listed in Table~\ref{tab:Delta}~\footnote{
    Although the effective pairing interaction generally contains off-diagonal components $V^{(\Gamma)}_{ij}$, these components vanish for the set of the basis pair potentials listed in Table~\ref{tab:Delta}.
}.
In the main text, the parameters of the tight-binding Hamiltonian in Eq.~\eqref{eq:wp} and the chemical potential $\mu$ are set to the same values as those used in Ref.~\onlinecite{yoda2026double}; these values are summarized in Table~\ref{tab:parameter}.
Here, we assume that Cooper pairs are formed predominantly by electrons near the Fermi surface.
Accordingly, the momentum integration in Eq.~\eqref{eq:gap_eq_matrix} is restricted to the region satisfying $|\xi_\ell(\bm{k})|,|\xi_{\ell'}(\bm{k})|<\omega_\mathrm{c}$.
In the numerical calculation, the cutoff energy is set to $\omega_\mathrm{c}=0.01$.
The bulk Brillouin zone shown in Fig.~\ref{fig:model_BZ}\subref{fig:BZ} is discretized using a uniform $125\times125\times125$ $\bm{k}$-mesh.
The normalized momentum sum in Eq.~\eqref{eq:gap_eq_matrix} is evaluated as
\begin{equation}
    \frac{1}{N}\sum_{\bm{k}}(\cdots)\ \longrightarrow\ \frac{1}{n_\mathrm{k}^3}\sum_{\bm{k}}(\cdots),
\end{equation}
with $n_\mathrm{k}=125$.
The gap equation is evaluated at the fixed temperature $\kb T=1.0\times10^{-4}$, rather than by determining $\Tc$ separately for each irrep.
At each interaction parameter point, we calculate the largest eigenvalue $\lambda^{(\Gamma)}_\mathrm{max}(T)$ for each irrep $\Gamma$ and identify the irrep with the largest value.
We regard eigenvalues satisfying $\lambda^{(\Gamma)}_\mathrm{max}(T)\leq\delta$ as numerically zero, with the tolerance set to $\delta=10^{-8}$.
When the largest eigenvalue among all irreps satisfies this criterion, the corresponding parameter point is classified as non-superconducting phase.
Otherwise, the irrep with the largest eigenvalue is assigned as the superconducting phase.
The resulting phase diagrams are shown in Figs.~\ref{fig:phase_attractive} and \ref{fig:phase_repulsive}, which correspond to attractive and repulsive on-site interactions, $V_\mathrm{on}=-0.1$ and $V_\mathrm{on}=0.1$, respectively.
We parametrize the off-site pairing interactions as
\begin{equation}
    V'_\parallel=\alpha_\parallel V_\parallel,\quad V'_\perp=\alpha_\perp V_\perp.
\end{equation}
In Figs.~\ref{fig:phase_attractive} and \ref{fig:phase_repulsive}, the columns correspond to the values of $\alpha_\parallel=0.0$, $0.5$, and $1.0$ from left to right, whereas the rows correspond to the values of $\alpha_\perp=0.0$, $0.5$, and $1.0$ from bottom to top.
In each phase diagram, the horizontal and vertical axes represent $V_\parallel$ and $V_\perp$, respectively, both of which vary from $-0.15$ to $0.15$.
Each interaction parameter, $V_\parallel$ and $V_\perp$, is sampled at 131 equally spaced points, corresponding to a grid spacing of $\varDelta V_\parallel=\varDelta V_\perp\simeq2.3\times10^{-3}$.

We first focus on the case in which $V'_\parallel=V'_\perp=0$ [Figs.~\ref{fig:phase_attractive}\subref{fig:attractive_0.0_0.0} and \ref{fig:phase_repulsive}\subref{fig:repulsive_0.0_0.0}].
These results show that in the regime with $V_\parallel<0$ and $V_\perp<0$, the $\mathrm{A_{1g}}$ pairing channel, which corresponds to $s$-wave superconductivity, is stabilized.
In addition, in part of the regime where $V_\parallel<0$ and $V_\perp>0$, the $\mathrm{A_{2u}}$ pairing is the leading channel.
This pairing supports double Majorana Kramers pairs at the $\M$ point, as discussed in Ref.~\onlinecite{yoda2026double}.
In the repulsive on-site case [Fig.~\ref{fig:phase_repulsive}\subref{fig:repulsive_0.0_0.0}], there is a non-superconducting phase in the regime for small $|V_\parallel|$ and $|V_\perp|$, and the $\mathrm{A_{2g}}$ pairing also appears in part of the region with $V_\parallel>0$ and $V_\perp<0$.
This pairing has symmetry-protected nodes in the bulk states along the $\bar{\Gamma}\M$ line and does not host Majorana Kramers pairs at the $\M$ point~\cite{yoda2026double}.
Of particular interest in the present work is the $\mathrm{A_{1u}}$ channel, which becomes dominant in the region with $V_\parallel>0$ and $V_\perp>0$, as well as in part of the region with $V_\parallel>0$ and $V_\perp<0$.
This indicates that the twisted surface spectrum discussed in Ref.~\onlinecite{yoda2026double} can be realized in these interaction regimes.

These features of the phase diagrams in Figs.~\ref{fig:phase_attractive}\subref{fig:attractive_0.0_0.0} and \ref{fig:phase_repulsive}\subref{fig:repulsive_0.0_0.0} can be qualitatively understood from the symmetry of the pair potential.
For instance, the pair potential $\hat{\Delta}_1^{(\mathrm{A_{1u}})}=s_0\rhoz\sigmaz$ has the opposite sign between in-plane nearest-neighbor sublattices [$\mathrm{A}(\mathrm{C})$ and $\mathrm{B}(\mathrm{D})$] and also between out-of-plane nearest-neighbor sublattices [$\mathrm{A}(\mathrm{B})$ and $\mathrm{C}(\mathrm{D})$].
This sign-changing structure is favored by positive pair-hopping amplitudes on the corresponding bonds, $V_\parallel>0$ and $V_\perp>0$, and is reflected in $V^{(\mathrm{A_{1u}})}_1=\frac{1}{4}V_\mathrm{on}-V_\parallel-\frac{1}{2}V_\perp$, explaining why the $\mathrm{A_{1u}}$ pairing is dominant in this region.
The same argument can apply to the other representations.
Nevertheless, the detailed structure of the phase diagram is not determined solely by effective pairing interactions $V^{(\Gamma)}_i$.
It also depends on the other factors such as the product of the form factor $\hat{\Delta}^{(\Gamma)}_i(\bm{k})$ and the projection operator $\mathcal{P}_\ell(\bm{k})$ [see Eq.~\eqref{eq:gap_eq_matrix}], and is therefore sensitive to the parameters of the normal Hamiltonian.

We next consider the effects of finite off-site density--density interactions, $V'_\parallel$ and $V'_\perp$.
For $V_\mathrm{on}=-0.1$ [Fig.~\ref{fig:phase_attractive}], increasing $\alpha_\perp$ stabilizes the $\mathrm{E_u}$ pairing in part of the region with $V_\parallel>0$ and $V_\perp<0$, regardless of the value of $\alpha_\parallel$ [Figs.~\ref{fig:phase_attractive}\subref{fig:attractive_0.0_1.0}--\subref{fig:attractive_1.0_1.0}].
For $V_{\mathrm{on}}=0.1$ [Fig.~\ref{fig:phase_repulsive}], the $\mathrm{E_u}$ phase is stabilized in the same region, but occupies a larger parameter range than for $V_\mathrm{on}=-0.1$ [Figs.~\ref{fig:phase_repulsive}\subref{fig:repulsive_0.0_1.0}--\subref{fig:attractive_1.0_0.5}].
In addition, as $\alpha_\parallel$ and $\alpha_\perp$ increase, the $\mathrm{B_{2u}}$ pairing becomes the leading channel near the origin on the $V_\parallel<0$ side, where both $|V_\parallel|$ and $|V_\perp|$ are small [Figs.~\ref{fig:phase_repulsive}\subref{fig:repulsive_0.5_1.0}, \subref{fig:repulsive_1.0_1.0}, \subref{fig:repulsive_0.5_0.5}, and \subref{fig:repulsive_1.0_0.5}].
To summarize, the off-site interactions, $V'_\parallel$ and $V'_\perp$, stabilize the off-site $\mathrm{B_{2u}}$ and $\mathrm{E_u}$ pairings, with the latter replacing the on-site $\mathrm{A_{1u}}$ pairing in part of the parameter region.
Nevertheless, a substantial part of the $\mathrm{A_{1u}}$ phase found in the absence of the off-site density--density interactions [Figs.~\ref{fig:phase_attractive}\subref{fig:attractive_0.0_0.0} and \ref{fig:phase_repulsive}\subref{fig:repulsive_0.0_0.0}] remains stable even in the presence of these interactions.


\begin{table}
    \begin{ruledtabular}
        \centering
        \caption{
            Compatibility relations connecting the irreps of the bulk point group $\mathrm{D_{4h}}$ to those of the surface point group $\mathrm{C_{4v}}$.
            The rightmost column lists the corresponding superconducting gap structures of the surface wallpaper fermions obtained in the effective surface theory~\cite{yoda2026superconducting}.
        }
        \begin{tabular}{ccc}
            PG $\mathrm{D_{4h}}$ & PG $\mathrm{C_{4v}}$ & Gap structure                               \\ \hline
            $\mathrm{B_{2u}}$    & $\mathrm{B_1}$       & Point nodes along the $\bar{\Gamma}\M$ line \\
            ${}^1\!\mathrm{E_u}$ & ${}^1\!\mathrm{E}$   & Line nodes except along the $\Y\M$ line     \\
        \end{tabular}
        \label{tab:sc_gap}
    \end{ruledtabular}
\end{table}
\begin{table}
    \begin{ruledtabular}
        \centering
        \caption{
            One-dimensional topological invariants defined on the $\mathrm{MA}$ line, provided that the bulk spectrum is fully gapped there, and the number of Majorana Kramers pairs (\#MKP) at the $\M$ point.
            We denote each order-two crystalline symmetry as $C_{2\mathrm{z}}\coloneq\{2_{001}|\bm{0}\}$, $M_\mathrm{x}\coloneq\{m_{100}|1/2\ 1/2\ 0\}$, $M_\mathrm{y}\coloneq\{m_{010}|1/2\ 1/2\ 0\}$, $M_\mathrm{D}\coloneq\{m_{110}|1/2\ 1/2\ 0\}$, and $M_{\bar{\mathrm{D}}}\coloneq\{m_{1\bar{1}0}|1/2\ 1/2\ 0\}$.
            Detailed calculations of these invariants are summarized in Appendix~\ref{app:1D_topo}.
        }
        \begin{tabular}{ccc}
            Irrep                & Topological invariants                                                        & \#MKP \\ \hline
            $\mathrm{B_{2u}}$    & $\nu^\pm_\mathrm{DIII}[M_\mathrm{x}]=\nu^\pm_\mathrm{DIII}[M_\mathrm{y}]=1$   & $2$   \\
                                 & $\nu^\pm_\mathrm{D}[M_\mathrm{D}]=\nu^\pm_\mathrm{D}[M_{\bar{\mathrm{D}}}]=0$ &       \\
                                 & $W[C_{2\mathrm{z}}]=0$                                                        &       \\ [5pt]
            ${}^1\!\mathrm{E_u}$ & $\nu^\pm_\mathrm{DIII}[M_\mathrm{y}]=1$                                       & $2$   \\
                                 & $\nu^\pm_\mathrm{D}[C_{2\mathrm{z}}]=0$                                       &       \\
        \end{tabular}
        \label{tab:topo}
    \end{ruledtabular}
\end{table}
\begin{figure*}
    \centering

    \subfloat[Normal state.]{
        \includegraphics[keepaspectratio,width=0.36\textwidth]
        {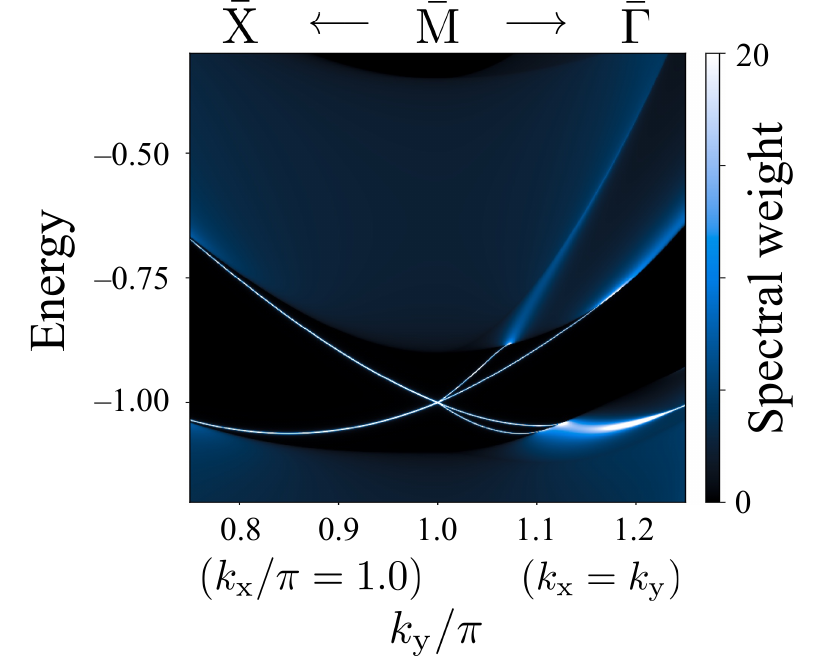}
        \label{fig:NS_ene}
    }\hfil
    \subfloat[$\mathrm{B_{2u}}$ representation.]{
        \includegraphics[keepaspectratio,width=0.35\textwidth]
        {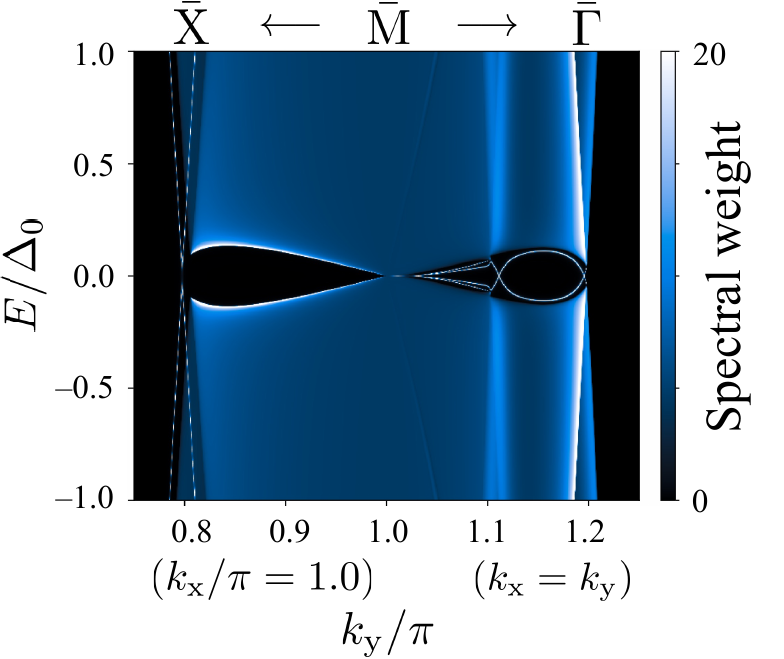}
        \label{fig:B2u_ene}
    }\\

    \subfloat[${}^1\!\mathrm{E_u}$ representation. ($\X\M\bar{\Gamma}$ path)]{
        \includegraphics[keepaspectratio,width=0.35\textwidth]
        {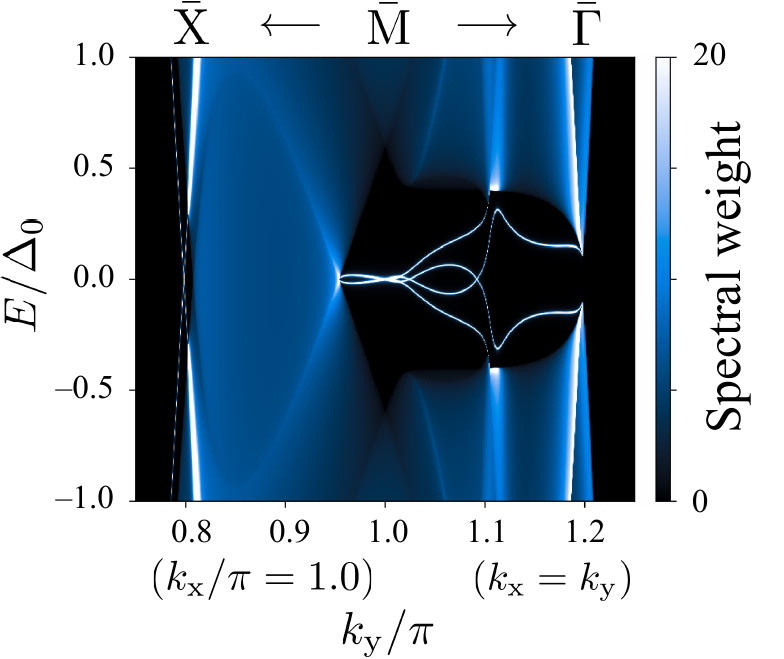}
        \label{fig:Eu_ene_X}
    }\hfil
    \subfloat[${}^1\!\mathrm{E_u}$ representation. ($\Y\M\bar{\Gamma}$ path)]{
        \includegraphics[keepaspectratio,width=0.35\textwidth]
        {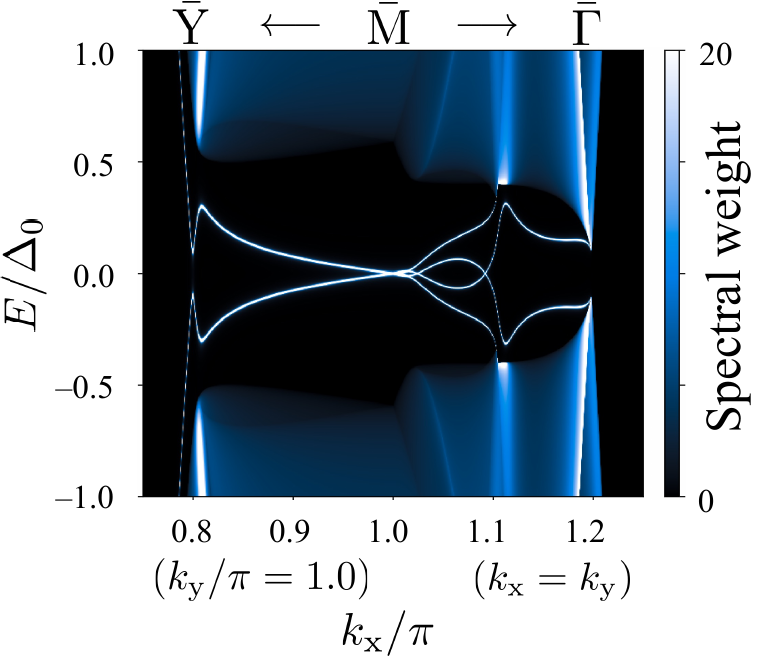}
        \label{fig:Eu_ene_Y}
    }

    \caption{(Color online)
        Surface spectral function for a semi-infinite system with the (001) surface on the C--D layer.
        The chemical potential and pairing amplitude are set to $\mu=-0.75$ and $\Delta_0=0.02$, respectively.
    }
    \label{fig:ene}
\end{figure*}

\section{Surface states with off-site pairings} \label{sec:surface_state}

In Sec.~\ref{sec:phase_result}, we showed that the finite off-site density--density interactions, $V'_\parallel$ and $V'_\perp$, stabilize the off-site $\mathrm{B_{2u}}$, and $\mathrm{E_u}$ pairing channels.
In this section, we focus on the symmetry-enforced features of the surface energy spectra characteristic of these pairing channels.
We first derive the spectral structures expected from crystalline symmetry and then verify them numerically.
Our primary objective is to determine how the gapless surface states are reconstructed by superconductivity, with particular emphasis on the hybridization between gapless wallpaper fermions and Majorana Kramers pairs.
Here, ``hybridization'' refers to a reconstruction of the surface energy spectrum resulting from the continuously connecting gapless surface branches. This includes both hybridization among the branches inherited from normal-state wallpaper fermions and hybridization between these branches and the spectra emanating from Majorana Kramers pairs pinned at the $\M$ point.

In the weak-coupling regime, in which the surface wallpaper fermions are assumed not to couple to the bulk states, whether the bulk pair potential gaps the surface spectrum inherited from the normal-state wallpaper fermions can be inferred from the effective surface theory developed in Ref.~\onlinecite{yoda2026superconducting}, which summarizes the relation between the surface pairing symmetry and the resulting superconducting gap structure for wallpaper-fermion spectra.
Combining this theory with the compatibility relations between irreps of the bulk point group $\mathrm{D_{4h}}$ and those of the surface point group $\mathrm{C_{4v}}$, we obtain the gap structures for surface wallpaper fermions summarized in Table~\ref{tab:sc_gap}.

The Majorana Kramers pairs at the $\M$ point can be regarded as edge modes of the one-dimensional topological crystalline superconductor along the $\mathrm{MA}$ line ($(\kx,\ky)=(\pi,\pi),\ -\pi\leq\kz\leq\pi$), characterized by one-dimensional topological invariants defined by order-two crystalline symmetries that leave this line invariant.
These invariants are determined by the space-group symmetry, the $\bm{k}$ point, and pairing symmetry~\cite{yamazaki2021magnetic}.
Applying the criteria of Ref.~\onlinecite{yamazaki2021magnetic} to the present model, we obtain the topological invariants and the number of Majorana Kramers pairs at the $\M$ point listed in Table~\ref{tab:topo}, under the assumption that the bulk spectrum is fully gapped along the $\mathrm{MA}$ line.
A detailed discussion is given in Appendix~\ref{app:1D_topo}.

In the following numerical analysis, we calculate the surface Green's function $G_\infty$ for the surface unit cell containing the $\mathrm{A}$, $\mathrm{B}$, $\mathrm{C}$, and $\mathrm{D}$ sublattices using the recursive Green's function method based on the M\"{o}bius transformation~\cite{umerski1997closed}.
The surface spectral function $A(\kx,\ky,E)$ is defined as
\begin{equation}
    A(\kx,\ky,E)\coloneq-\frac{1}{\pi}\Im\Tr G_\infty(\kx,\ky,E+i\delta),
    \label{eq:spectral}
\end{equation}
where $E$ is the energy and $\delta>0$ is a small broadening parameter, set to $\delta=1.0\times10^{-4}$ in the normal state and $\delta=5.0\times10^{-6}$ in the superconducting state.
For the C--D layer, the trace in Eq.~\eqref{eq:spectral} is taken over the spin degrees of freedom on the C and D sublattices in the normal state, while both particle and hole sectors are additionally included in the superconducting state.
For the momentum resolution, the $\X-\M$, $\Y-\M$, and $\M-\bar{\Gamma}$ directions are each divided into 300 intervals, from $(\pi,\frac{3}{4}\pi)$ to $(\pi,\pi)$, from $(\frac{3}{4}\pi,\pi)$ to $(\pi,\pi)$, and from $(\pi,\pi)$ to $(\frac{5}{4}\pi,\frac{5}{4}\pi)$, respectively.
For the energy resolution, the range $-1.2\leq E\leq-0.3$ in the normal state and $-\Delta_0\leq E\leq\Delta_0$ in the superconducting state are both divided into 600 intervals.
Here, $\Delta_0$ is a pairing amplitude and is fixed to $\Delta_0=0.02$.
No normalization is applied to the spectral function, and the color scale is fixed to $0\leq A\leq 20$, with values above 20 saturated at the upper limit.
Figure~\ref{fig:ene}\subref{fig:NS_ene} shows the surface energy spectrum of the normal states along the $\X\M\bar{\Gamma}$ path.


\subsection{\texorpdfstring{$\mathrm{B_{2u}}$}{B2u} channel} \label{sec:B2u}
First, we examine the surface spectrum for the $\mathrm{B_{2u}}$ channel with $\Delta_0\hat{\Delta}^{(\mathrm{B_{2u}})}_1(\bm{k})$.

\paragraph*{Theoretical analysis.}
The $\mathrm{B_{2u}}$ representation of the bulk point group $\mathrm{D_{4h}}$ is reduced to the $\mathrm{B_1}$ representation of the surface point group $\mathrm{C_{4v}}$, whose pair potential produces symmetry-protected point nodes in the wallpaper-fermion spectrum along the $\bar{\Gamma}\M$ lines~\cite{yoda2026superconducting}.
Thus, the wallpaper fermions remain gapless along these lines in the $\mathrm{B_{2u}}$ representation.
We next consider the possible emergence of Majorana Kramers pairs at the $\M$ point.
Provided that the bulk BdG spectrum is fully gapped along the $\mathrm{MA}$ line, a $\Z_2$ topological invariant $\nu^\pm_\mathrm{DIII}[M_\mathrm{x}]$ associated with $M_\mathrm{x}\coloneq\{m_{100}|1/2\ 1/2\ 0\}$ can be defined in each $\pm 1$ eigenspace of this glide operation.
In the weak-coupling limit, this invariant is determined by the Fermi-surface criterion~\cite{yamazaki2021magnetic},
\begin{equation}
    \nu^\pm_\mathrm{DIII}[M_\mathrm{x}]=\frac{\#\FS}{4} \pmod 2,
    \label{eq:nu_DIII}
\end{equation}
where $\#\FS$ denotes the number of Fermi-level crossings, including degeneracy, along the half $\mathrm{MA}$ line from $\mathrm{M}$ ($\kz=0$) to $\mathrm{A}$ ($\kz=\pi$).
For the parameter set considered here, $\#\FS=4$, and hence we obtain $\nu^\pm_\mathrm{DIII}[M_\mathrm{x}]=1$.
The nontrivial invariants in each glide eigenspace imply the emergence of double Majorana Kramers pairs at the $\M$ point.
The same Fermi-surface criterion gives $\nu^\pm_\mathrm{DIII}[M_\mathrm{y}]=1$ for $M_\mathrm{y}\coloneq\{m_{010}|1/2\ 1/2\ 0\}$.
Since $M_\mathrm{y}$ anticommutes with $M_\mathrm{x}$, however, $\nu^\pm_\mathrm{DIII}[M_\mathrm{y}]$ is not independent of $\nu^\pm_\mathrm{DIII}[M_\mathrm{x}]$.
Therefore, when the bulk spectrum is gapped along the $\mathrm{MA}$ line, the $\mathrm{B_{2u}}$ pairing state is expected to support the coexistence of the gapless wallpaper fermions and double Majorana Kramers pairs along the $\bar{\Gamma}\M$ line.

\paragraph*{Numerical results.}
Figure~\ref{fig:ene}\subref{fig:B2u_ene} shows the surface spectral function for the nearest-neighbor $\mathrm{B_{2u}}$ pair potential.
The wallpaper fermions crossing the Fermi level in the normal state become gapped along the $\X\M$ line, whereas they remain gapless along the $\bar{\Gamma}\M$ line.
This behavior is consistent with the effective surface theory~\cite{yoda2026superconducting} summarized in Table~\ref{tab:sc_gap}.
Along the $\bar{\Gamma}\M$ line, the surface energy spectrum is reconstructed and exhibits the twisted structure owing to hybridization between gapless wallpaper fermions themselves in the region $\kx=\ky,\ 1.1\lesssim\ky/\pi\lesssim1.2$.

However, the one-dimensional topological invariants discussed above cannot be directly applied to the present model since $\hat{\Delta}^{(\mathrm{B_{2u}})}(\pi,\pi,\kz)=0$ [see the explicit form of $\hat{\Delta}^{(\mathrm{B_{2u}})}_1(\bm{k})$ in Table~\ref{tab:Delta}], and the bulk BdG spectrum consequently remains gapless along the $\mathrm{MA}$ line.
Thus, Majorana Kramers pairs cannot be defined at the $\M$ point.
According to the Mackey--Bradley theorem~\cite{mackey1953symmetric, bradley1970kronecker, bradley2009mathematical}, this bulk gap closing is accidental rather than symmetry-enforced; it results from restricting the $\mathrm{B_{2u}}$ pair potential to the in-plane nearest-neighbor component.
Therefore, within nearest-neighbor pairings, the $\mathrm{B_{2u}}$ pairing exhibits the twisted spectrum arising from hybridization between the gapless wallpaper fermions themselves.
The surface-spectrum result including the symmetry-allowed longer-range $\mathrm{B_{2u}}$ pairing components is provided in Appendix~\ref{app:B2u}.


\begin{table}
    \begin{ruledtabular}
        \centering
        \caption{
            Solutions of the linearized gap equation~\eqref{eq:gap_eq} for the ${}^1\!\mathrm{E_u}$ representation corresponding to the largest eigenvalue.
            The eigenvectors are normalized as $\abs*{{}^1\!\bm{\eta}^{(\mathrm{E_u})}}=1$.
        }
        \begin{tabular}{cr}
            Expansion coefficients ${}^1\!\bm{\eta}^{(\mathrm{E_u})}$ & \multicolumn{1}{c}{Solution of Eq.~\eqref{eq:gap_eq}} \\ \hline
            ${}^1\!\eta^{(\mathrm{E_u})}_1$                           & $3.865879632033\times10^{-1}$                         \\
            ${}^1\!\eta^{(\mathrm{E_u})}_2$                           & $1.403245198559\times10^{-1}$                         \\
            ${}^1\!\eta^{(\mathrm{E_u})}_3$                           & $3.120328670753\times10^{-1}$                         \\
            ${}^1\!\eta^{(\mathrm{E_u})}_4$                           & $-5.986556823275\times10^{-2}$                        \\
            ${}^1\!\eta^{(\mathrm{E_u})}_5$                           & $-7.984107700756\times10^{-3}$                        \\
            ${}^1\!\eta^{(\mathrm{E_u})}_6$                           & $-7.452682438551\times10^{-1}$                        \\
            ${}^1\!\eta^{(\mathrm{E_u})}_7$                           & $2.234736756246\times10^{-1}$                         \\
            ${}^1\!\eta^{(\mathrm{E_u})}_8$                           & $-3.528192093220\times10^{-1}$                        \\
        \end{tabular}
        \label{tab:eta}
    \end{ruledtabular}
\end{table}

\subsection{\texorpdfstring{$\mathrm{E_u}$}{Eu} channel} \label{sec:Eu}

We turn to the surface spectrum of a representative time-reversal-invariant nematic $\mathrm{E_u}$ state.
We choose its first component, denoted by ${}^1\!\mathrm{E_u}$, while the second component describes the symmetry-related nematic orientation obtained by a fourfold rotation.
The pair potential is written as ${}^1\!\hat{\Delta}^{(\mathrm{E_u})}(\bm{k})\coloneq\Delta_0\sum_{i=1}^{8} {}^1\!\eta^{\mathrm{(E_u)}}_i {}^1\!\hat{\Delta}^{(\mathrm{E_u})}_i(\bm{k})$.
We determine the expansion coefficients ${}^1\!\bm{\eta}^{(\mathrm{E_u})}$ from the normalized eigenvector corresponding to the largest eigenvalue $\lambda_\mathrm{max}^{(\Gamma)}$ of Eq.~\eqref{eq:gap_eq} at $\kb T=1.0\times10^{-4}$, which is the fixed temperature used in Sec.~\ref{sec:phase_result}.
The interaction parameters are taken to be the same as those used for the phase diagram shown in Fig.~\ref{fig:phase_repulsive}\subref{fig:repulsive_1.0_1.0}, namely $V_\mathrm{on}=0.1$ and $V'_\parallel=V_\parallel$, $V'_\perp=V_\perp$.
We choose a point in the $\mathrm{E_u}$ phase, $V_\parallel=2.7\times10^{-2}$ and $V_\perp=-9.0\times10^{-3}$ in Fig.~\ref{fig:phase_repulsive}\subref{fig:repulsive_1.0_1.0}, and the resulting eigenvector is summarized in Table~\ref{tab:eta}.

\paragraph*{Theoretical analysis.}
The ${}^1\!\mathrm{E_u}$ representation is compatible with the ${}^1\!\mathrm{E}$ representation, which is the first component of the $\mathrm{E}$ representation of the point group $\mathrm{C_{4v}}$.
According to the effective surface theory~\cite{yoda2026superconducting}, in the ${}^1\!\mathrm{E_u}$ pairing state, the wallpaper fermions remain gapless except along the $\Y\M$ line.
Furthermore, following the criteria derived in Ref.~\onlinecite{yamazaki2021magnetic}, we obtain nontrivial topological invariants, $\nu^\pm_\mathrm{DIII}[M_\mathrm{y}]=1$, which indicates the presence of double Majorana Kramers pairs at the $\M$ point.
Thus, the ${}^1\!\mathrm{E_u}$ pairing state supports the coexistence of gapless wallpaper fermions and double Majorana Kramers pairs, except along the $\Y\M$ line where the wallpaper fermion spectrum is gapped.

\paragraph*{Numerical results.}
Figures~\ref{fig:ene}\subref{fig:Eu_ene_X} and \subref{fig:Eu_ene_Y} show the surface spectral functions for the ${}^1\!\mathrm{E_u}$ pairing along the $\X\M\bar{\Gamma}$ and $\Y\M\bar{\Gamma}$ paths, respectively.
Along the $\X\M$ line, the wallpaper-fermion spectrum remains gapless, whereas a superconducting gap opens along the $\Y\M$ line.
This anisotropic gap structure is consistent with the effective surface theory~\cite{yoda2026superconducting}.
In addition, double Majorana Kramers pairs protected by $\nu^\pm_\mathrm{DIII}[M_\mathrm{y}]=1$ appear at the $\M$ point.
Although Majorana Kramers pairs coexist with gapless wallpaper fermions along the $\X\M$ line, their hybridization does not occur because of crystalline-symmetry-protected bulk nodes on this line.

Along the $\bar{\Gamma}\M$ line, the gapless wallpaper fermion branch at $(\kx/\pi,\ky/\pi)\simeq(1.1,1.1)$ hybridizes with the branches originating from the Majorana Kramers pairs at the $\M$ point, giving rise to a characteristic twisted surface spectrum along the $\kx=\ky$, $1.0\lesssim\ky/\pi\lesssim1.1$.
By contrast, a superconducting gap opens at $(\kx/\pi,\ky/\pi)\simeq(1.2,1.2)$ in apparent disagreement with the effective surface theory summarized in Table~\ref{tab:sc_gap}.
We attribute this discrepancy to coupling between the surface wallpaper fermions and the projected bulk states for the present parameter set.
As shown in Appendix~\ref{app:Eu}, the expected gapless branches become visible when the chemical potential $\mu$ is adjusted to separate the surface states from the bulk continuum.

\section{Discussion} \label{sec:discussion}

In this section, we combine the phase-diagram results in Sec.~\ref{sec:phase_result} with the surface-state analysis in Sec.~\ref{sec:surface_state}, to clarify whether the characteristic surface spectra persist in the presence of off-site interactions.
When only the on-site pairing channels are considered, the $\mathrm{A_{1u}}$ phase occupies a substantial part of the region with a positive pair-hopping amplitude $V_\parallel>0$, where it gives rise to the twisted surface spectrum discussed in Ref.~\onlinecite{yoda2026double}.
The off-site pairing interactions $V'_\parallel$ and $V'_\perp$, however, yield a regime in which the off-site $\mathrm{B_{2u}}$ and $\mathrm{E_u}$ phases are stabilized and, especially, the on-site $\mathrm{A_{1u}}$ phase is partly replaced by the $\mathrm{E_u}$ phase.
Importantly, the suppression of the $\mathrm{A_{1u}}$ phase does not necessarily eliminate the characteristic hybridized surface spectra because the representative time-reversal-invariant nematic $\mathrm{E_u}$ state considered here also supports both gapless wallpaper fermions and Majorana Kramers pairs along the $\bar{\Gamma}\M$ line and exhibits the hybridized surface spectral structure.
Furthermore, in the $\mathrm{B_{2u}}$ pairing, gapless wallpaper fermions remain along the $\bar{\Gamma}\M$ line and form a twisted surface spectrum through their hybridization.
These results indicate that twisted surface structures originating from hybridization phenomena are expected over a wide range of interaction parameters even in the presence of the off-site pairing interactions.

\section{Conclusion} \label{sec:conclusion}

In this work, we theoretically investigated the pairing phase diagram and the superconducting surface spectra of a topological nonsymmorphic crystalline insulator with wallpaper fermions, taking into account both on-site and off-site pairing channels.
When the pair potential is restricted to the on-site channels, the $\mathrm{A_{1u}}$ pairing state, which possesses the twisted surface spectrum, becomes the leading channel over a substantial region with a positive in-plane nearest-neighbor pair-hopping amplitude $V_\parallel>0$.
This tendency can be understood from the sign-changing structure of the $\mathrm{A_{1u}}$ pair potential between nearest-neighbor sublattices.
The off-site density--density interactions, $V'_\parallel$ and $V'_\perp$, stabilize the $\mathrm{E_u}$ pairing in part of the region with $V_\parallel>0$, regardless of the sign of the on-site interaction $V_\mathrm{on}$.
Moreover, for $V_\mathrm{on}>0$, the off-site $\mathrm{B_{2u}}$ pairing becomes the leading channel in the region with $V_\parallel<0$ and small $|V_\parallel|$ and $|V_\perp|$.
Nevertheless, the on-site $\mathrm{A_{1u}}$ pairing remains the leading channel over a substantial parameter region even in the presence of these off-site interactions.

We also clarified the surface spectral structures of the pairing states induced by the off-site interactions.
For the nearest-neighbor $\mathrm{B_{2u}}$ pairing considered in the main text, the wallpaper fermions remain gapless along the $\bar{\Gamma}\M$ line and form a characteristic twisted surface spectrum through hybridization between the remaining wallpaper-fermion branches themselves.
For this nearest-neighbor pairing model, accidental bulk nodes along the $\mathrm{MA}$ line render the relevant one-dimensional topological invariants and the corresponding Majorana Kramers pairs at the $\M$ point ill-defined.
As shown in Appendix~\ref{app:B2u}, symmetry-allowed longer-range $\mathrm{B_{2u}}$ components gap these nodes and allow double Majorana Kramers pairs to emerge.
The representative time-reversal-invariant nematic $\mathrm{E_u}$ state considered here, chosen as the ${}^1\!\mathrm{E_u}$ component, exhibits an anisotropic surface gap structure and supports both gapless wallpaper fermions and double Majorana Kramers pairs.
Along the $\X\M$ line, symmetry-protected bulk nodes prevent hybridization between these modes.
Along the $\bar{\Gamma}\M$ line, by contrast, the wallpaper-fermion branches hybridize with branches emanating from the double Majorana Kramers pairs and form a twisted surface spectrum.
Thus, even when off-site interactions alter the leading pairing channel, hybridization-driven surface-spectrum reconstruction can persist in the nearest-neighbor $\mathrm{B_{2u}}$ state and in the representative time-reversal-invariant nematic $\mathrm{E_u}$ state analyzed here.

\acknowledgments
This work is supported by JST SPRING (Grant No.~JPMJSP2125) and JSPS KAKENHI (Grant Nos.~JP24H00853 and JP25K07224).

\section*{Data availability}
The data that support the findings of this article will be publicly available~\cite{data}.

\appendix


\begin{table}
    \begin{ruledtabular}
        \centering
        \caption{
            Symmetry operators $U$ and $P$ on the $\mathrm{MA}$ line in the space group $P4/mbm$.
        }
        \begin{tabular}{cl}
                & \multicolumn{1}{c}{Symmetry operator}                                           \\ \hline
            $U$ & $\{2_{001}|\bm{0}\}$, $\{m_{100}|1/2\ 1/2\ 0\}$, $\{m_{010}|1/2\ 1/2\ 0\}$      \\
                & $\{m_{110}|1/2\ 1/2\ 0\}$, $\{m_{1\bar{1}0}|1/2\ 1/2\ 0\}$                      \\ [5pt]
            $P$ & $\{2_{100}|1/2\ 1/2\ 0\}$, $\{2_{010}|1/2\ 1/2\ 0\}$, $\{2_{110}|1/2\ 1/2\ 0\}$ \\
                & $\{2_{1\bar{1}0}|1/2\ 1/2\ 0\}$, $\{m_{001}|\bm{0}\}$, $\{I|\bm{0}\}$           \\
        \end{tabular}
        \label{tab:order-two}
    \end{ruledtabular}
\end{table}

\section{One-dimensional topological invariants on the MA lines} \label{app:1D_topo}

In this appendix, we summarize the criteria derived in Ref.~\onlinecite{yamazaki2021magnetic} for one-dimensional topological invariants associated with order-two crystalline symmetries and apply them to the superconducting states considered in the main text.
For the Fourier transformation used here, we adopt a convention different from that in Eq.~\eqref{eq:fourier} of the main text:
\begin{equation}
    c^\dag_{\bm{k},X,s}\coloneq\frac{1}{\sqrt{N}}\sum_{\bm{n}}e^{i\bm{k}\cdot\bm{n}}c^\dag_{\bm{n}+\bm{\tau}_X,s},
    \label{eq:fourier_app}
\end{equation}
where the sublattice-dependent phase factors $e^{i\bm{k}\cdot\bm{\tau}_X}$ are not included.
With this convention, the Hamiltonian is written in a periodic gauge such that $H_\BdG(\bm{k})=H_\BdG(\bm{k}+\bm{G})$, where $\bm{G}$ is a reciprocal lattice vector, and hence the Hamiltonian can be identified across the Brillouin-zone boundaries.
Throughout this appendix, our discussion is restricted to the $\mathrm{MA}$ line and we write the momentum as $\bm{k}=(\bm{k}_\pi,\kz)$, where $\bm{k}_\pi$ denotes the momentum parallel to the surface $(\kx,\ky)=(\pi,\pi)$.
The order-two symmetries play two distinct roles, denoted by $U$ and $P$.
The symmetry $U$ leaves each momentum on the $\mathrm{MA}$ line invariant and thus, satisfies
\begin{equation}
    [H_\wp(\bm{k}_\pi,\kz),D_{\bm{k}}(U)]=0,
\end{equation}
whereas $P$ reverses the momentum on the $\mathrm{MA}$ line such that
\begin{equation}
    D_{\bm{k}}(P)H_\wp(\bm{k}_\pi,\kz)D_{\bm{k}}(P)^\dag=H_\wp(\bm{k}_\pi,-\kz).
    \label{eq:def_P}
\end{equation}
Here, $D_{\bm{k}}(g)$ denotes the representation of the symmetry operation $g$ for normal Hamiltonian at momentum $\bm{k}$.
For the $\mathrm{MA}$ line in the space group $P4/mbm$, $U$ and $P$ are taken to be those listed in Table~\ref{tab:order-two}.
We further define the parity $\chi(g)$ of the pair potential under the symmetry $g$ by
\begin{equation}
    D_{\bm{k}}(g)\hat{\Delta}^{(\Gamma)}(\bm{k})D_{\bm{k}}(g)^\dagger=\chi(g)\hat{\Delta}^{(\Gamma)}(g\bm{k}).
\end{equation}
The type of one-dimensional topological invariants that can be defined is determined by the values of $\chi(U)$ and $D_{\bm{k}}(U)^2$, while $P$ plays an auxiliary role in the establishment of the criteria for these invariants and their evaluation~\cite{yamazaki2021magnetic}.

For type B (magnetic chiral) in Table~II in Ref.~\onlinecite{yamazaki2021magnetic}, that is $\chi(U)=1$ and $D_{\bm{k}}(U)^2=-1$, a magnetic winding number $W[U]\in\Z$ can take a nonzero value.
Its explicit form is given by~\cite{xiong2017anisotropica}
\begin{equation}
    W[U]\coloneq\frac{i}{4\pi}\int_{-\pi}^{\pi}\dd{\kz}\Tr[\Gamma[U]H^{-1}_\BdG(\bm{k})\pdv{H_\BdG(\bm{k})}{\kz}],
    \label{eq:mag_winding}
\end{equation}
where we use the magnetic chiral operator $\Gamma[U]\coloneq e^{i\phi}(D_{\bm{k}}(U)\tau_0)\tauy$, instead of the ordinary chiral operator $\Gamma\coloneq\tauy$.
Here, $D_{\bm{k}}(U)\tau_0$ denotes the representation of $U$ in Nambu space and the phase $\phi$ is chosen such that $\Gamma[U]^2=1$.
In the present wallpaper fermion system, because of four-dimensional irreps at the $\M$ point in the wallpaper group $p4g$~\cite{Aroyo2011-cr, Aroyo2006-bi1, Aroyo2006-bi2, elcoro2021magnetic, xu2020high}, the magnetic winding number in Eq.~\eqref{eq:mag_winding} is quantized to $W[U]\in4\Z$.
Furthermore, using other criteria derived in Ref.~\onlinecite{xiong2017anisotropica} associated with additional order-two crystalline symmetries, we can diagnose whether $W[U]$ takes a finite value.
One of them is that $W[U]$ is restricted to zero when there exists a symmetry $P$ such that $\{D_{\bm{k}}(U),D_{\bm{k}}(P)\}=0$ and $\chi(P)=-1$.
We can choose $P=\{2_{100}|1/2\ 1/2\ 0\}$ for the magnetic winding numbers $W[C_{2\mathrm{z}}]$ for the $\mathrm{B_{2u}}$ representation and the invariant is therefore forced to vanish as shown in Table~\ref{tab:topo}.

In the case of $\chi(U)=D_{\bm{k}}(U)^2=-1$, which corresponds to type C (symmorphic) in Table~II in Ref.~\onlinecite{yamazaki2021magnetic}, we can define two equivalent one-dimensional topological invariants in class D, $\nu^+_\mathrm{D}[U]=\nu^-_\mathrm{D}[U]\in\Z_2$, in each eigenspace of $U$.
On the $\mathrm{MA}$ line in the space group $P4/mbm$, $\nu^\pm_\mathrm{D}[U]$ with $U=M_\mathrm{D}$ and $M_{\bar{\mathrm{D}}}$ ($C_{2\mathrm{z}}$) can be introduced in the $\mathrm{B_{2u}}$ (${}^1\!\mathrm{E_u}$) representation [see Table~\ref{tab:topo}].
Here, we can choose $I\coloneq\{I|\bm{0}\}$ as the symmetry $P$ satisfying both Eq.~\eqref{eq:def_P} and
\begin{equation}
    [D_{\bm{k}}(U),D_{\bm{k}}(P)]=0,\quad D_{\bm{k}}(P)^2=1.
    \label{eq:"inversion"}
\end{equation}
Such a symmetry $P$ is regarded as the ``inversion'' symmetry along the $\mathrm{MA}$ line in each eigensector of $U$.
Therefore, the Bloch state and the state obtained by applying $I\Theta$ to that Bloch state belong to the different eigenspaces and are degenerate for arbitrary $\bm{k}$.
Since $\chi(I)=-1$ for the $\mathrm{B_{2u}}$ and ${}^1\!\mathrm{E_u}$ representations, following the procedure shown in Sec.~VB2 in Ref.~\onlinecite{yamazaki2021magnetic}, topological invariants $\nu^\pm_\mathrm{D}[U]$ in the weak-coupling limit can be calculated using the Fermi-surface criterion given by
\begin{equation}
    \nu^\pm_\mathrm{D}[U]=\frac{\#\FS}{2} \pmod 2,
\end{equation}
where $\#\FS$ is defined in the same manner as in the main text.
In addition, the $P4/mbm$ symmetry enforces fourfold-degenerate bands along this line~\cite{Aroyo2011-cr, Aroyo2006-bi1, Aroyo2006-bi2, elcoro2021magnetic, xu2020high} and consequently, $\#\FS$ is constrained to be a multiple of four, leading to $\nu^\pm_\mathrm{D}[U]=0$.

Finally, when $\chi(U)=D_{\bm{k}}(U)^2=1$, which corresponds to type D (nonsymmorphic) in Table~II in Ref.~\onlinecite{yamazaki2021magnetic}, two independent one-dimensional $\Z_2$ topological invariants in class DIII, $\nu^+_\mathrm{DIII}[U]$ and $\nu^-_\mathrm{DIII}[U]$, can be defined in each eigenspace of $U$.
As an example, we take $U=M_\mathrm{y}\coloneq\{m_{010}|1/2\ 1/2\ 0\}$.
$M_\mathrm{y}$ anticommutes with $M_\mathrm{x}\coloneq\{m_{100}|1/2\ 1/2\ 0\}$, implying that $\nu^+_\mathrm{DIII}[M_\mathrm{y}]=\nu^-_\mathrm{DIII}[M_\mathrm{y}]$.
Furthermore, along the $\mathrm{MA}$ line, we can choose $\{I|\bm{0}\}$ as the symmetry $P$ satisfying both Eqs.~\eqref{eq:def_P} and \eqref{eq:"inversion"}.
Therefore, according to the criterion shown in Fig.~5 in Ref.~\onlinecite{yamazaki2021magnetic}, Fermi-surface criterion in Eq.~\eqref{eq:nu_DIII} is applicable when the pair potential is odd under spatial inversion $\{I|\bm{0}\}$.
For the normal-state parameter set used in the main text, $\#\FS=4$, and hence we get $\nu^\pm_{\mathrm{DIII}}[M_\mathrm{y}]=1$.
For a fully gapped $\mathrm{B_{2u}}$ state, the same argument also applies to $U=M_\mathrm{x}$.
As mentioned in Sec.~\ref{sec:B2u}, even when both $\nu^\pm_{\mathrm{DIII}}[M_\mathrm{x}]$ and $\nu^\pm_{\mathrm{DIII}}[M_\mathrm{y}]$ can be defined and nontrivial, they cannot be regarded as independent topological invariants due to $[M_\mathrm{x},M_\mathrm{y}]\neq0$~\footnote{
    For the case of $U=M_\mathrm{x}$ ($M_\mathrm{y}$), the ``inversion'' operator $P$ can also be chosen as $P=\{2_{100}|1/2\ 1/2\ 0\}$ ($\{2_{010}|1/2\ 1/2\ 0\}$).
}.

\begin{figure*}
    \centering

    \subfloat[Normal state.]{
        \includegraphics[keepaspectratio,width=0.325\textwidth]
        {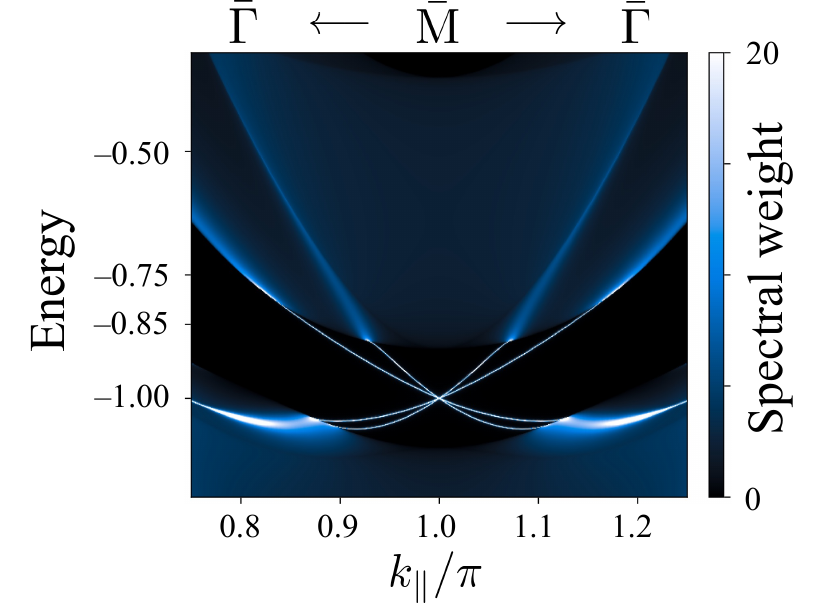}
        \label{fig:NS_ene_app}
    }\hfil
    \subfloat[$\mathrm{B_{2u}}$ representation.]{
        \includegraphics[keepaspectratio,width=0.31\textwidth]
        {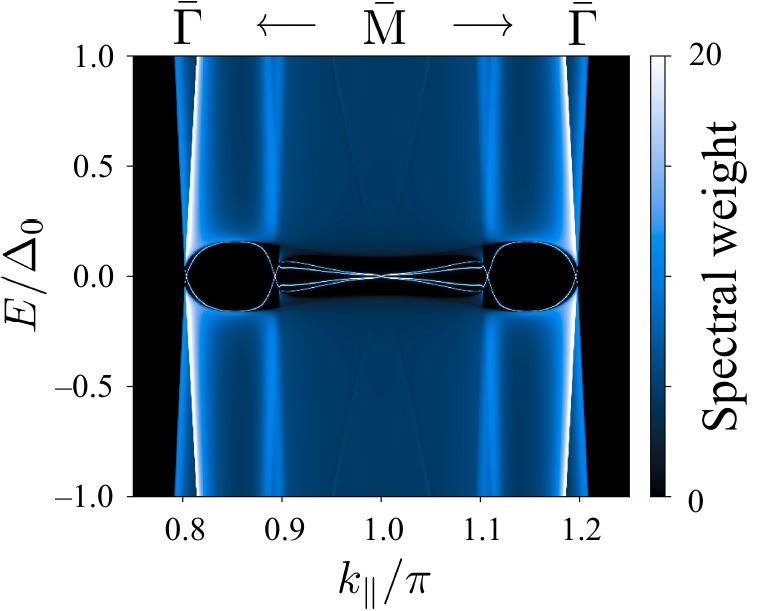}
        \label{fig:B2u_ene_app}
    }
    \hfil
    \subfloat[${}^1\!\mathrm{E_u}$ representation.]{
        \includegraphics[keepaspectratio,width=0.31\textwidth]
        {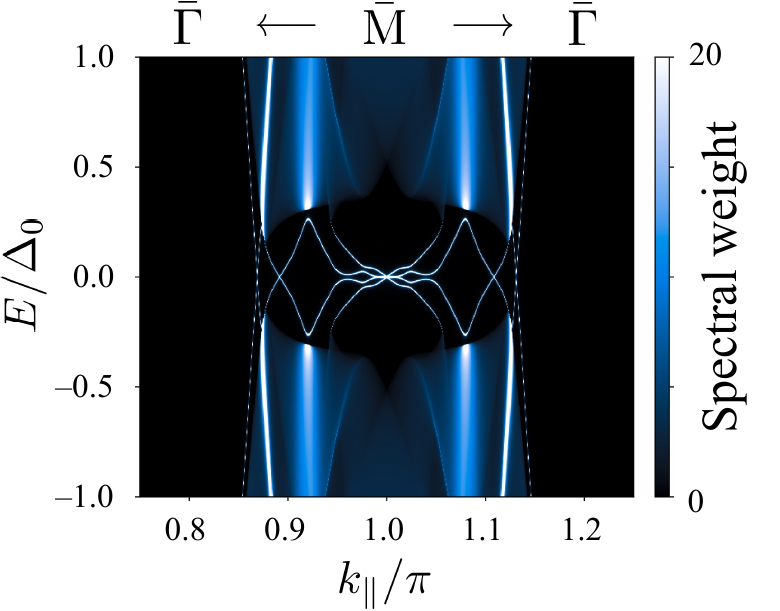}
        \label{fig:Eu_ene_app}
    }

    \caption{(Color online)
        Surface spectral functions for a semi-infinite system with the (001) surface on the C--D layer along the $\bar{\Gamma}\M\bar{\Gamma}$ line.
        The parameters of the tight-binding Hamiltonian and the pairing amplitude $\Delta_0$ are the same as those used in the main text.
        The chemical potential is set to (b) $\mu=-0.75$ and (c) $\mu=-0.85$.
    }
    \label{fig:ene_app}
\end{figure*}

\section{Supplementary details on the surface energy spectra} \label{app:ene}


In this appendix, we provide additional results for the surface energy spectra in the $\mathrm{B_{2u}}$ and ${}^1\!\mathrm{E_u}$ pairing states.
We focus on the spectrum along the $\bar{\Gamma}\M\bar{\Gamma}$ line and introduce the surface momentum $k_\parallel$ along this line, defined as $(\kx,\ky)=(k_\parallel,k_\parallel)$.
In the following calculation, the parameter set for the tight-binding Hamiltonian is the same as that listed in Table~\ref{tab:parameter} and the pairing amplitude is fixed to $\Delta_0=0.02$.
Figure~\ref{fig:ene_app}\subref{fig:NS_ene_app} shows the surface spectral function in the normal state along the $\bar{\Gamma}\M\bar{\Gamma}$ line.

\subsection{\texorpdfstring{$\mathrm{B_{2u}}$}{B2u} channel} \label{app:B2u}
The pair potential vanishes on the $\mathrm{MA}$ line, producing bulk nodes at the normal-state Fermi crossings; these nodes project onto the surface $\M$ point.
We then include the symmetry-allowed out-of-plane next-nearest-neighbor pairings [$\mathrm{A(B)}$ and $\mathrm{D(C)}$] as follows:
\begin{align}
     & \hat{\Delta}^{(\mathrm{B_{2u}})}_2(\bm{k})\coloneq\sin{\frac{\kx}{2}}\sin{\frac{\ky}{2}}\sin{\frac{\kz}{2}}s_0\rhox\sigmay, \label{eq:B2u_2} \\
     & \hat{\Delta}^{(\mathrm{B_{2u}})}_3(\bm{k})\coloneq\sin{\frac{\kx}{2}}\sin{\frac{\ky}{2}}\sin{\frac{\kz}{2}}\sz\rhox\sigmax. \label{eq:B2u_3}
\end{align}
The pair potential used in the calculation is $\hat{\Delta}^{(\mathrm{B_{2u}})}(\bm{k})\coloneq\Delta_0\sum_{i=1}^3\eta^{(\mathrm{B_{2u}})}_i\hat{\Delta}^{(\mathrm{B_{2u}})}_i(\bm{k})$ with $\bm{\eta}^{(\mathrm{B_{2u}})}=(2/\sqrt{5},1/\sqrt{10},1/\sqrt{10})$.
Figure~\ref{fig:ene_app}\subref{fig:B2u_ene_app} shows the resulting surface spectral function along the $\bar{\Gamma}\M\bar{\Gamma}$ line with $\mu=-0.75$.
The longer-range pairing components in Eqs.~\eqref{eq:B2u_2} and \eqref{eq:B2u_3} open a gap in the accidental projected bulk spectrum at the $\M$ point.
Once such components open a bulk gap, the Fermi-surface criterion in Eq.~\eqref{eq:nu_DIII} becomes applicable.
Because the normal-state Fermi-surface configuration remains unchanged, it gives $\nu^\pm_\mathrm{DIII}[M_\mathrm{x}]=\nu^\pm_\mathrm{DIII}[M_\mathrm{y}]=1$.
Consequently, double Majorana Kramers pairs protected by $\nu^\pm_\mathrm{DIII}[M_\mathrm{x}]=1$ or $\nu^\pm_\mathrm{DIII}[M_\mathrm{y}]=1$ appear at this point.
The newly emerging double Majorana Kramers pairs hybridize with crystalline-symmetry-protected gapless wallpaper fermions, producing the twisted surface spectrum along the $\bar{\Gamma}\M$ line.

\subsection{\texorpdfstring{${}^1\!\mathrm{E_u}$}{1Eu} channel} \label{app:Eu}
In Sec.~\ref{sec:Eu}, we pointed out that, in the ${}^1\!\mathrm{E_u}$ pairing with $\mu=-0.75$, the wallpaper-fermion branch near $k_\parallel/\pi\simeq1.2$ is gapped, contrary to the prediction of the effective surface theory~\cite{yoda2026superconducting}.
This apparent gap originates from mixing between the surface branches and the projected bulk continuum.
To separate these states, we shift the chemical potential to $\mu=-0.85$.
The resulting surface spectral function is shown in Fig.~\ref{fig:ene_app}\subref{fig:Eu_ene_app}.
At this chemical potential, the wallpaper fermion branch near $k_\parallel/\pi\simeq1.2$ remains gapless, in agreement with the effective surface theory.
Together with the gapless branch near $k_\parallel/\pi\simeq1.1$, they hybridize with the branches emanating from the double Majorana Kramers pairs at the $\M$ point, giving rise to the characteristic twisted surface spectrum.

\bibliography{ref}

@article{hao2011surface,
  title = {Surface Spectral Function in the Superconducting State of a Topological Insulator},
  author = {Hao, Lei and Lee, T. K.},
  year = {2011},
  month = apr,
  journal = {Phys. Rev. B},
  volume = {83},
  numpages = {13},
  pages = {134516},
  publisher = {American Physical Society},
  doi = {10.1103/PhysRevB.83.134516},
  urldate = {2024-09-23}
}

@article{hashimoto2015surface,
  title = {Surface Electronic State of Superconducting Topological Crystalline Insulator},
  author = {Hashimoto, Tatsuki and Yada, Keiji and Sato, Masatoshi and Tanaka, Yukio},
  year = {2015},
  month = nov,
  journal = {Phys. Rev. B},
  volume = {92},
  numpages = {17},
  pages = {174527},
  publisher = {American Physical Society},
  doi = {10.1103/PhysRevB.92.174527},
  urldate = {2024-09-22}
}

@article{hsieh2012Majorana,
  title = {Majorana {Fermions} and {Exotic} {Surface} {Andreev} {Bound} {States} in {Topological} {Superconductors}: {Application} to {Cu}${}_x${Bi}\textsubscript{2}{Se}\textsubscript{3}},
  author = {Hsieh, Timothy H. and Fu, Liang},
  journal = {Phys. Rev. Lett.},
  volume = {108},
  issue = {10},
  pages = {107005},
  numpages = {5},
  year = {2012},
  month = {Mar},
  publisher = {American Physical Society},
  doi = {10.1103/PhysRevLett.108.107005}
}

@article{kawakami2018topological,
  title = {Topological {{Crystalline Materials}} of ${J}=3/2$ {{Electrons}}: {{Antiperovskites}}, {{Dirac Points}}, and {{High Winding Topological Superconductivity}}},
  shorttitle = {Topological {{Crystalline Materials}} of {{J}} = 3 / 2 {{Electrons}}},
  author = {Kawakami, Takuto and Okamura, Tetsuya and Kobayashi, Shingo and Sato, Masatoshi},
  year = {2018},
  month = nov,
  journal = {Phys. Rev. X},
  volume = {8},
  numpages = {4},
  pages = {041026},
  issn = {2160-3308},
  doi = {10.1103/PhysRevX.8.041026},
  urldate = {2024-09-13},
  langid = {english}
}

@article{lu2015crossed,
  title = {Crossed {{Surface Flat Bands}} of {{Weyl Semimetal Superconductors}}},
  author = {Lu, Bo and Yada, Keiji and Sato, Masatoshi and Tanaka, Yukio},
  year = {2015},
  month = mar,
  journal = {Phys. Rev. Lett.},
  volume = {114},
  numpages = {9},
  pages = {096804},
  issn = {0031-9007, 1079-7114},
  doi = {10.1103/PhysRevLett.114.096804},
  urldate = {2025-08-19},
  copyright = {http://link.aps.org/licenses/aps-default-license},
  langid = {english}
}

@article{mizuno2023hall,
  title = {Hall Effect of Ferro/Antiferromagnetic Wallpaper Fermions},
  author = {Mizuno, Koki and Yamakage, Ai},
  year = {2023},
  month = jun,
  journal = {Phys. Rev. B},
  volume = {107},
  numpages = {23},
  pages = {235301},
  publisher = {American Physical Society},
  doi = {10.1103/PhysRevB.107.235301},
  urldate = {2024-09-23}
}

@article{novak2013unusual,
  title = {Unusual Nature of Fully Gapped Superconductivity in {{In-doped SnTe}}},
  author = {Novak, Mario and Sasaki, Satoshi and Kriener, Markus and Segawa, Kouji and Ando, Yoichi},
  year = {2013},
  month = oct,
  journal = {Phys. Rev. B},
  volume = {88},
  numpages = {14},
  pages = {140502(R)},
  doi = {10.1103/PhysRevB.88.140502}
}

@article{sigrist1991phenomenological,
  title = {Phenomenological theory of unconventional superconductivity},
  author = {Sigrist, Manfred and Ueda, Kazuo},
  year = {1991},
  month = apr,
  journal = {Rev. Mod. Phys.},
  volume = {63},
  number = {2},
  pages = {239--311},
  issn = {0034-6861, 1539-0756},
  doi = {10.1103/RevModPhys.63.239},
  urldate = {2024-09-13},
  copyright = {http://link.aps.org/licenses/aps-default-license},
  langid = {english}
}

@article{wieder2018wallpaper,
  title = {Wallpaper Fermions and the Nonsymmorphic {{Dirac}} Insulator},
  author = {Wieder, Benjamin J. and Bradlyn, Barry and Wang, Zhijun and Cano, Jennifer and Kim, Youngkuk and Kim, Hyeong-Seok D. and Rappe, Andrew M. and Kane, C. L. and Bernevig, B. Andrei},
  year = {2018},
  month = jul,
  journal = {Science},
  volume = {361},
  number = {6399},
  pages = {246--251},
  publisher = {American Association for the Advancement of Science},
  doi = {10.1126/science.aan2802},
  urldate = {2025-06-02}
}

@article{yamakage2012theory,
  title = {Theory of Tunneling Conductance and Surface-State Transition in Superconducting Topological Insulators},
  author = {Yamakage, Ai and Yada, Keiji and Sato, Masatoshi and Tanaka, Yukio},
  year = {2012},
  month = may,
  journal = {Phys. Rev. B},
  volume = {85},
  numpages = {18},
  pages = {180509(R)},
  issn = {1098-0121, 1550-235X},
  doi = {10.1103/PhysRevB.85.180509},
  urldate = {2024-09-13},
  copyright = {http://link.aps.org/licenses/aps-default-license},
  langid = {english}
}

@article{yamakage2013theory,
  title = {Theory of Tunneling Spectroscopy in a Superconducting Topological Insulator},
  author = {Yamakage, Ai and Yada, Keiji and Sato, Masatoshi and Tanaka, Yukio},
  year = {2013},
  month = nov,
  journal = {Physica C: Superconductivity},
  volume = {494},
  pages = {20--23},
  issn = {09214534},
  doi = {10.1016/j.physc.2013.04.019},
  urldate = {2024-10-02},
  langid = {english}
}

@article{yamazaki2021magnetic,
  title = {Magnetic Response of {{Majorana Kramers}} Pairs with an Order-Two Symmetry},
  author = {Yamazaki, Yuki and Kobayashi, Shingo and Yamakage, Ai},
  year = {2021},
  month = mar,
  journal = {Phys. Rev. B},
  volume = {103},
  numpages = {9},
  pages = {094508},
  publisher = {American Physical Society},
  doi = {10.1103/PhysRevB.103.094508},
}

@article{mackey1953symmetric,
  author = {George W. Mackey},
  journal = {Am. J. Math.},
  number = {2},
  pages = {387},
  title = {Symmetric and {Anti} {Symmetric} {Kronecker} {Squares} and {Intertwining} {Numbers} of {Induced} {Representations} of {Finite} {Groups}},
  urldate = {2025-08-19},
  volume = {75},
  year = {1953},
  doi = {10.2307/2372459}
}

@article{bradley1970kronecker,
  author = {Bradley, C. J. and Davies, B. L.},
  journal = {J. Math. Phys.},
  number = {5},
  pages = {1536},
  title = {Kronecker {Products} and {Symmetrized} {Squares} of {Irreducible} {Representations} of {Space} {Groups}},
  urldate = {2025-08-19},
  volume = {11},
  year = {1970},
  doi = {10.1063/1.1665292}
}

@article{zhou2021glide,
  title = {Glide Symmetry Protected Higher-Order Topological Insulators from Semimetals with Butterfly-like Nodal Lines},
  author = {Zhou, Xiaoting and Hsu, Chuang-Han and Huang, Cheng-Yi and Iraola, Mikel and Ma{\~n}es, Juan L. and Vergniory, Maia G. and Lin, Hsin and Kioussis, Nicholas},
  year = {2021},
  month = dec,
  journal = {npj Comput. Mater.},
  volume = {7},
  number = {1},
  pages = {202},
  publisher = {Nature Publishing Group},
  issn = {2057-3960},
  doi = {10.1038/s41524-021-00672-9},
  urldate = {2025-08-19},
  copyright = {2021 The Author(s)},
  langid = {english}
}

@article{hwang2023magnetic,
  title = {Magnetic Wallpaper {{Dirac}} Fermions and Topological Magnetic {{Dirac}} Insulators},
  author = {Hwang, Yoonseok and Qian, Yuting and Kang, Junha and Lee, Jehyun and Ryu, Dongchoon and Choi, Hong Chul and Yang, Bohm-Jung},
  year = {2023},
  month = apr,
  journal = {npj Comput. Mater.},
  volume = {9},
  number = {1},
  pages = {65},
  issn = {2057-3960},
  doi = {10.1038/s41524-023-01018-3}
}

@article{ryu2020wallpaper,
  title = {Wallpaper {{Dirac Fermion}} in a {{Nonsymmorphic Topological Kondo Insulator}}: {Pu}{B}${}_4$},
  author = {Ryu, Dong-Choon and Kim, Junwon and Choi, Hongchul and Min, Byung Il},
  year = {2020},
  month = nov,
  journal = {J. Am. Chem. Soc.},
  volume = {142},
  number = {45},
  pages = {19278--19282},
  publisher = {American Chemical Society},
  issn = {0002-7863},
  doi = {10.1021/jacs.0c09442}
}

@article{mizuno2025magnon,
  title = {Magnon-mediated superconductivity at the interface between a ferromagnetic insulator and a topological crystalline insulator with wallpaper fermions},
  author = {Mizuno, Koki and Yamakage, Ai},
  journal = {Phys. Rev. B},
  volume = {112},
  issue = {3},
  pages = {035303},
  numpages = {11},
  year = {2025},
  month = {Jul},
  publisher = {American Physical Society},
  doi = {10.1103/sknv-r7wq}
}

@book{bradley2009mathematical,
  title = {The {Mathematical} {Theory} {Of} {Symmetry} {In} {Solids}: {Representation} theory for point groups and space groups},
  isbn = {978-0-19-958258-7},
  publisher = {Oxford University Press},
  author = {Bradley, C J and Cracknell, A P},
  month = dec,
  year = {2009},
  doi = {10.1093/oso/9780199582587.001.0001}
}

@article{Aroyo2011-cr,
	author = {Aroyo, M. I. and Perez-Mato, J. M. and Orobengoa, D. and Tasci, E. and De La Flor, G. and Kirov, A.},
	title = {Crystallography online: Bilbao crystallographic server},
	year = {2011},
	journal = {Bulgarian Chemical Communications},
	volume = {43},
	number = {2},
	pages = {183--197}
}

@article{Aroyo2006-bi1,
  title       = {Bilbao {Crystallographic} {Server}: I. {Databases} and crystallographic computing programs},
  author      = {Mois Ilia Aroyo and Juan Manuel Perez-Mato and Cesar Capillas and Eli Kroumova and Svetoslav Ivantchev and Gotzon Madariaga and Asen Kirov and Hans Wondratschek},
  pages       = {15--27},
  volume      = {221},
  number      = {1},
  journal     = {Zeitschrift f\"{u}r Kristallographie - Crystalline Materials},
  doi         = {10.1524/zkri.2006.221.1.15},
  year        = {2006}
}

@article{Aroyo2006-bi2,
  author   = {Aroyo, Mois I. and Kirov, Asen and Capillas, Cesar and Perez-Mato, J. M. and Wondratschek, Hans},
  title    = {{Bilbao Crystallographic Server. II. Representations of crystallographic point groups and space groups}},
  journal  = {Acta Cryst. A},
  year     = {2006},
  volume   = {62},
  number   = {2},
  pages    = {115--128},
  month    = {Mar},
  doi      = {10.1107/S0108767305040286}
}

@article{elcoro2021magnetic,
  title={Magnetic topological quantum chemistry},
  author={Elcoro, Luis and Wieder, Benjamin J and Song, Zhida and Xu, Yuanfeng and Bradlyn, Barry and Bernevig, B Andrei},
  journal={Nat. Commun.},
  volume={12},
  number={1},
  pages={5965},
  year={2021},
  doi= {10.1038/s41467-021-26241-8},
  publisher={Nature Publishing Group UK London}
}

@article{xu2020high,
  title={High-throughput calculations of magnetic topological materials},
  author={Xu, Yuanfeng and Elcoro, Luis and Song, Zhi-Da and Wieder, Benjamin J and Vergniory, Maia G and Regnault, Nicolas and Chen, Yulin and Felser, Claudia and Bernevig, B Andrei},
  journal={Nature},
  volume={586},
  number={7831},
  pages={702--707},
  year={2020},
  doi={10.1038/s41586-020-2837-0},
  publisher={Nature Publishing Group UK London}
}

@article{Aroyo2014brillouin,
  author = "Aroyo, Mois I. and Orobengoa, Danel and de la Flor, Gemma and Tasci, Emre S. and Perez-Mato, J. Manuel and Wondratschek, Hans",
  title = "{Brillouin-zone database on the Bilbao Crystallographic Server}",
  journal = "Acta Cryst. A",
  year = "2014",
  volume = "70",
  number = "2",
  pages = "126--137",
  month = "Mar",
  doi = {10.1107/S205327331303091X},
}

@article{yonezawa2018nematic,
  title={Nematic {Superconductivity} in {Doped} {Bi}\textsubscript{2}{Se}\textsubscript{3} {Topological} {Superconductors}},
  author={Yonezawa, Shingo},
  journal={Condens. Matter},
  volume={4},
  number={1},
  pages={2},
  year={2018},
  publisher={MDPI},
  doi={10.3390/condmat4010002}
}

@article{sasaki2012odd,
  title = {Odd-{Parity} {Pairing} and {Topological} {Superconductivity} in a {Strongly} {Spin-Orbit} {Coupled} {Semiconductor}},
  author = {Sasaki, Satoshi and Ren, Zhi and Taskin, A. A. and Segawa, Kouji and Fu, Liang and Ando, Yoichi},
  journal = {Phys. Rev. Lett.},
  volume = {109},
  issue = {21},
  pages = {217004},
  numpages = {5},
  year = {2012},
  month = {Nov},
  publisher = {American Physical Society},
  doi = {10.1103/PhysRevLett.109.217004}
}

@article{wang2016observation,
  title={Observation of superconductivity induced by a point contact on 3{D} {Dirac} semimetal {Cd}\textsubscript{3}{As}\textsubscript{2} crystals},
  author={Wang, He and Wang, Huichao and Liu, Haiwen and Lu, Hong and Yang, Wuhao and Jia, Shuang and Liu, Xiong-Jun and Xie, XC and Wei, Jian and Wang, Jian},
  journal={Nat. Mater.},
  volume={15},
  number={1},
  pages={38--42},
  year={2016},
  publisher={Nature Publishing Group UK London},
  doi={10.1038/nmat4456}
}

@article{aggarwal2016unconventional,
  title={Unconventional superconductivity at mesoscopic point contacts on the 3{D} {Dirac} semimetal {Cd}\textsubscript{3}{As}\textsubscript{2}},
  author={Aggarwal, Leena and Gaurav, Abhishek and Thakur, Gohil S and Haque, Zeba and Ganguli, Ashok K and Sheet, Goutam},
  journal={Nat. Mater.},
  volume={15},
  number={1},
  pages={32--37},
  year={2016},
  publisher={Nature Publishing Group UK London},
  doi={10.1038/nmat4455}
}

@article{wang2017discovery,
  title={Discovery of tip induced unconventional superconductivity on {Weyl} semimetal},
  author={Wang, He and Wang, Huichao and Chen, Yuqin and Luo, Jiawei and Yuan, Zhujun and Liu, Jun and Wang, Yong and Jia, Shuang and Liu, Xiong-Jun and Wei, Jian and Wang, Jian},
  journal={Sci. Bulletin},
  volume={62},
  number={6},
  pages={425--430},
  year={2017},
  publisher={Elsevier},
  doi={10.1016/j.scib.2017.02.009}
}

@article{he2013full,
  title = {Full superconducting gap in the doped topological crystalline insulator {Sn}${}_{0.6}${In}${}_{0.4}${Te}},
  author = {He, L. P. and Zhang, Z. and Pan, J. and Hong, X. C. and Zhou, S. Y. and Li, S. Y.},
  journal = {Phys. Rev. B},
  volume = {88},
  issue = {1},
  pages = {014523},
  numpages = {4},
  year = {2013},
  month = {Jul},
  publisher = {American Physical Society},
  doi = {10.1103/PhysRevB.88.014523}
}

@article{kobayashi2015topological,
  title = {Topological {Superconductivity} in {Dirac} {Semimetals}},
  author = {Kobayashi, Shingo and Sato, Masatoshi},
  journal = {Phys. Rev. Lett.},
  volume = {115},
  issue = {18},
  pages = {187001},
  numpages = {5},
  year = {2015},
  month = {Oct},
  publisher = {American Physical Society},
  doi = {10.1103/PhysRevLett.115.187001}
}

@article{umerski1997closed,
  title = {Closed-form solutions to surface {Green's} functions},
  author = {Umerski, A.},
  journal = {Phys. Rev. B},
  volume = {55},
  issue = {8},
  pages = {5266--5275},
  numpages = {0},
  year = {1997},
  month = {Feb},
  publisher = {American Physical Society},
  doi = {10.1103/PhysRevB.55.5266}
}

@article{xiong2017anisotropica,
  title = {Anisotropic {{Magnetic Responses}} of {{Topological Crystalline Superconductors}}},
  author = {Xiong, Yuansen and Yamakage, Ai and Kobayashi, Shingo and Sato, Masatoshi and Tanaka, Yukio},
  year = {2017},
  month = feb,
  journal = {Crystals},
  volume = {7},
  number = {2},
  pages = {58},
  publisher = {Multidisciplinary Digital Publishing Institute},
  issn = {2073-4352},
  doi = {10.3390/cryst7020058},
  }

@article{schnyder2008classification,
  title = {Classification of topological insulators and superconductors in three spatial dimensions},
  author = {Schnyder, Andreas P. and Ryu, Shinsei and Furusaki, Akira and Ludwig, Andreas W. W.},
  journal = {Phys. Rev. B},
  volume = {78},
  issue = {19},
  pages = {195125},
  numpages = {22},
  year = {2008},
  month = {Nov},
  publisher = {American Physical Society},
  doi = {10.1103/PhysRevB.78.195125},
}

@article{ryu2010topological,
  title={Topological insulators and superconductors: tenfold way and dimensional hierarchy},
  author={Ryu, Shinsei and Schnyder, Andreas P and Furusaki, Akira and Ludwig, Andreas WW},
  journal={New J. Phys.},
  volume={12},
  number={6},
  pages={065010},
  year={2010},
  publisher={IOP Publishing},
  doi={10.1088/1367-2630/12/6/065010}
}

@article{kitaev2009periodic,
  title={Periodic table for topological insulators and superconductors},
  author={Kitaev, Alexei},
  journal={AIP Conf. Proc.},
  volume={1134},
  number={1},
  pages={22--30},
  year={2009},
  organization={American Institute of Physics},
  doi = {10.1063/1.3149495}
}

@article{mao2022third,
  title={Third-order topological insulators with wallpaper fermions in {Tl}\textsubscript{4}{Pb}{Te}\textsubscript{3} and {Tl}\textsubscript{4}{Sn}{Te}\textsubscript{3}},
  author={Mao, Ning and Wang, Hao and Dai, Ying and Huang, Baibiao and Niu, Chengwang},
  journal={npj Comput. Mater.},
  volume={8},
  number={1},
  pages={154},
  year={2022},
  doi = {10.1038/s41524-022-00839-y}
}

@article{qi2011topological,
  title = {Topological insulators and superconductors},
  author = {Qi, Xiao-Liang and Zhang, Shou-Cheng},
  journal = {Rev. Mod. Phys.},
  volume = {83},
  issue = {4},
  pages = {1057--1110},
  numpages = {0},
  year = {2011},
  month = {Oct},
  publisher = {American Physical Society},
  doi = {10.1103/RevModPhys.83.1057},
}

@article{ivanov2001non-abelian,
  title = {Non-{Abelian} {Statistics} of {Half}-{Quantum} {Vortices} in $p$-{Wave} {Superconductors}},
  author = {Ivanov, D. A.},
  journal = {Phys. Rev. Lett.},
  volume = {86},
  issue = {2},
  pages = {268--271},
  numpages = {0},
  year = {2001},
  month = {Jan},
  publisher = {American Physical Society},
  doi = {10.1103/PhysRevLett.86.268},
}

@article{kitaev2006anyons,
  title = {Anyons in an exactly solved model and beyond},
  author = {Kitaev, Alexei},
  journal = {Ann. Phys.},
  volume = {321},
  issue = {1},
  pages = {2--111},
  year = {2006},
  doi = {10.1016/j.aop.2005.10.005},
}

@article{nayak2008non-abelian,
  title = {Non-{Abelian} anyons and topological quantum computation},
  author = {Nayak, Chetan and Simon, Steven H. and Stern, Ady and Freedman, Michael and Das Sarma, Sankar},
  journal = {Rev. Mod. Phys.},
  volume = {80},
  issue = {3},
  pages = {1083--1159},
  numpages = {0},
  year = {2008},
  month = {Sep},
  publisher = {American Physical Society},
  doi = {10.1103/RevModPhys.80.1083},
}

@article{yoda2026superconducting,
  title = {Superconducting {Gap} {Structures} in {Wallpaper} {Fermion} {Systems}},
  author = {Yoda, Kaito and Yamakage, Ai},
  journal = {J. Low Temp. Phys.},
  volume = {222},
  issue = {2},
  pages = {44},
  year = {2026},
  month = {Feb},
  doi = {10.1007/s10909-026-03368-w},
}

@article{sato-ando2017topological,
doi = {10.1088/1361-6633/aa6ac7},
year = {2017},
volume = {80},
number = {7},
pages = {076501},
author = {Sato, Masatoshi and Ando, Yoichi},
title = {Topological superconductors: a review},
journal = {Rep. Prog. Phys.},
}

@article{chiu2016classification,
  title = {Classification of topological quantum matter with symmetries},
  author = {Chiu, Ching-Kai and Teo, Jeffrey C. Y. and Schnyder, Andreas P. and Ryu, Shinsei},
  journal = {Rev. Mod. Phys.},
  volume = {88},
  issue = {3},
  pages = {035005},
  numpages = {63},
  year = {2016},
  month = {Aug},
  publisher = {American Physical Society},
  doi = {10.1103/RevModPhys.88.035005},
}

@article{beenakker2013search,
   author = "Beenakker, C.W.J.",
   title = "Search for {Majorana} {Fermions} in {Superconductors}", 
   journal= "Annu. Rev. Condens. Matter Phys.",
   year = "2013",
   volume = "4",
   number = "Volume 4, 2013",
   pages = "113-136",
   doi = "10.1146/annurev-conmatphys-030212-184337",
   publisher = "Annual Reviews",
  }

@article{sato2016majorana,
author = {Sato ,Masatoshi and Fujimoto ,Satoshi},
title = {Majorana {Fermions} and {Topology} in {Superconductors}},
journal = {J. Phys. Soc. Jpn.},
volume = {85},
number = {7},
pages = {072001},
year = {2016},
doi = {10.7566/JPSJ.85.072001}
}

@article{alicea2012new,
doi = {10.1088/0034-4885/75/7/076501},
year = {2012},
month = {jun},
publisher = {IOP Publishing},
volume = {75},
number = {7},
pages = {076501},
author = {Alicea, Jason},
title = {New directions in the pursuit of {Majorana} fermions in solid state systems},
journal = {Rep. Prog. Phys.},
}

@article{aguado2017majorana,
  title={Majorana quasiparticles in condensed matter},
  author={Aguado, Ram{\'o}n},
  journal={Riv. Nuovo Cim.},
  volume={40},
  number={11},
  pages={523--593},
  year={2017},
  publisher={Springer},
  doi={10.1393/ncr/i2017-10141-9}
}

@article{mizushima2016symmetry,
author = {Mizushima ,Takeshi and Tsutsumi ,Yasumasa and Kawakami ,Takuto and Sato ,Masatoshi and Ichioka ,Masanori and Machida ,Kazushige},
title = {Symmetry-{Protected} {Topological} {Superfluids} and {Superconductors} ---{From} the {Basics} to ${}^3${He}---},
journal = {J. Phys. Soc. Jpn.},
volume = {85},
number = {2},
pages = {022001},
year = {2016},
doi = {10.7566/JPSJ.85.022001},
}

@article{yoda2026double,
  title = {Double-twisted surface spectrum from hybridized {Majorana} {Kramers} pairs and wallpaper fermions},
  author = {Yoda, Kaito and Yamakage, Ai},
  journal = {Phys. Rev. B},
  volume = {113},
  issue = {22},
  pages = {224515},
  numpages = {9},
  year = {2026},
  month = {Jun},
  publisher = {American Physical Society},
  doi = {10.1103/vnfd-ngb3}
}

@article{nomoto2016classification,
  title = {Classification of ``multipole'' superconductivity in multiorbital systems and its implications},
  author = {Nomoto, T. and Hattori, K. and Ikeda, H.},
  journal = {Phys. Rev. B},
  volume = {94},
  issue = {17},
  pages = {174513},
  numpages = {16},
  year = {2016},
  month = {Nov},
  publisher = {American Physical Society},
  doi = {10.1103/PhysRevB.94.174513}
}

@article{tanaka2024theory,
  title={Theory of {Majorana} {Zero} {Modes} in {Unconventional} {Superconductors}},
  author={Tanaka, Yukio and Tamura, Shun and Cayao, Jorge},
  journal={Prog. Theor. Exp. Phys.},
  volume={2024},
  number={8},
  pages={08C105},
  year={2024},
  publisher={Oxford University Press},
  doi = {10.1093/ptep/ptae065}
}

@misc{data,
author = {Yoda, Kaito and Yamakage, Ai},
title = {Competition between on-site and off-site pairing phases and surface spectra in superconducting wallpaper fermion systems},
howpublished = {NAGOYA Repository},
note={\url{https://nagoya.repo.nii.ac.jp}}
}

\end{document}